\documentclass[longauth]{aa} 
 
\usepackage{graphicx}
 \usepackage[]{natbib}
\usepackage{txfonts}
\usepackage{booktabs}
\usepackage{url}
\usepackage{xcolor}
\usepackage{sidecap}
\usepackage{amssymb}
\bibpunct{(}{)}{;}{a}{}{,} 
\begin{document}

\title{The Gaia-ESO Survey: Exploring the complex nature and origins of the Galactic bulge populations\thanks{Based on data products from observations made with ESO Telescopes at the La Silla Paranal Observatory under programme ID 188.B-3002. These data products have been processed by the Cambridge Astronomy Survey Unit (CASU) at the Institute of Astronomy, University of Cambridge, and by the FLAMES/UVES reduction team at INAF/Osservatorio Astrofisico di Arcetri. These data have been obtained from the Gaia-ESO Survey Data Archive, prepared and hosted by the Wide Field Astronomy Unit, Institute for Astronomy, University of Edinburgh, which is funded by the UK Science and Technology Facilities Council.}}

    
\author{A.~Rojas-Arriagada \inst{\ref{inst1},\ref{inst5},\ref{inst6}}
  \and A.~Recio-Blanco \inst{\ref{inst1}}
  \and P.~de Laverny \inst{\ref{inst1}}
  \and \v{S}.~Mikolaitis \inst{\ref{inst10}}
  \and F.~Matteucci \inst{\ref{inst2},\ref{inst3},\ref{inst4}}
  \and E.~Spitoni  \inst{\ref{inst2},\ref{inst3}}
  \and M. Schultheis \inst{\ref{inst1}}
  \and M.~Hayden  \inst{\ref{inst1}}
  \and V. Hill \inst{\ref{inst1}}
  \and M. Zoccali \inst{\ref{inst5},\ref{inst6}} 
  \and D.~Minniti \inst{\ref{inst6},\ref{inst7},\ref{inst8}}
  \and O.~A.~Gonzalez\inst{\ref{inst9},\ref{inst11}}
  \and G.~Gilmore \inst{\ref{inst12}}
  \and S.~Randich \inst{\ref{inst13}}
  \and S.~Feltzing \inst{\ref{inst17}}
  \and E.~J.~Alfaro \inst{\ref{inst19}}
  \and C.~Babusiaux \inst{\ref{inst20}}
  \and T.~Bensby \inst{\ref{inst17}}
  \and A.~Bragaglia \inst{\ref{inst18}}
  \and E.~Flaccomio \inst{\ref{inst16}}
  \and S.~E.~Koposov \inst{\ref{inst12}}
  \and E.~Pancino \inst{\ref{inst13},\ref{inst21}}
  \and A.~Bayo \inst{\ref{inst25}}
  \and G.~Carraro \inst{\ref{inst9}}
  \and A.~R.~Casey \inst{\ref{inst12}}
  \and M.~T.~Costado \inst{\ref{inst19}}
  \and F.~Damiani \inst{\ref{inst16}}
  \and P.~Donati \inst{\ref{inst18}}
  \and E.~Franciosini \inst{\ref{inst13}}
  \and A.~Hourihane \inst{\ref{inst12}}
  \and P.~Jofr\'e \inst{\ref{inst12},\ref{inst24}}
  \and C.~Lardo \inst{\ref{inst15}}
  \and J.~Lewis \inst{\ref{inst12}}
  \and K.~Lind \inst{\ref{inst22},\ref{inst23}}
  \and L.~Magrini \inst{\ref{inst13}}
  \and L.~Morbidelli \inst{\ref{inst13}}
  \and G.~G.~Sacco \inst{\ref{inst13}}
  \and C.~C.~Worley \inst{\ref{inst12}}
  \and S.~Zaggia \inst{\ref{inst14}}
  }

\institute{
Laboratoire Lagrange, Universit\'e C\^ote d'Azur, Observatoire de la C\^ote d'Azur, CNRS, Bd de l'Observatoire, CS 34229, 06304 Nice cedex 4, France \email{arojas@oca.eu} \label{inst1}
\and
Instituto de Astrof\'{i}sica, Facultad de F\'{i}sica, Pontificia Universidad Cat\'olica de Chile, Av. Vicu\~na Mackenna 4860, Santiago, Chile \label{inst5}
\and
Millennium Institute of Astrophysics, Av. Vicu\~{n}a Mackenna 4860, 782-0436 Macul, Santiago, Chile \label{inst6}
\and
Institute of Theoretical Physics and Astronomy, Vilnius University, A. Go\v{s}tauto 12, 01108 Vilnius, Lithuania \label{inst10}
\and
Dipartimento di Fisica, Sezione di Astronomia, Universit\`a di Trieste, Via G.B. Tiepolo 11, I-34143 Trieste, Italy \label{inst2}
\and
INAF, Osservatorio Astronomico di Trieste, Via G.B. Tiepolo 11, I-34143 Trieste, Italy \label{inst3}
\and
INFN, Sezione di Trieste, Via A. Valerio 2 I-34127 Trieste, Italy \label{inst4}
\and
Departamento de Ciencias F\'isicas, Universidad Andr\'es Bello, 220 Rep\'ublica, Santiago, Chile \label{inst7}
\and
Vatican Observatory, V00120 Vatican City State, Italy \label{inst8}
\and
European Southern Observatory, Alonso de Cordova 3107 Vitacura, Santiago de Chile, Chile  \label{inst9}
\and
Institute for Astronomy, University of Edinburgh, Royal Observatory, Blackford Hill, Edinburgh, EH9 3HJ, UK \label{inst11}
\and
Institute of Astronomy, University of Cambridge, Madingley Road, Cambridge CB3 0HA, United Kingdom \label{inst12}
\and 
INAF - Osservatorio Astrofisico di Arcetri, Largo E. Fermi 5, 50125, Florence, Italy \label{inst13}
\and
Lund Observatory, Department of Astronomy and Theoretical Physics, Box 43, SE-221 00 Lund, Sweden \label{inst17}
\and
Instituto de Astrof\'{i}sica de Andaluc\'{i}a-CSIC, Apdo. 3004, 18080, Granada, Spain \label{inst19}
\and 
GEPI, Observatoire de Paris, CNRS, Universit\'e Paris Diderot, 5 Place Jules Janssen, 92190 Meudon, France \label{inst20}
\and
INAF - Osservatorio Astronomico di Bologna, via Ranzani 1, 40127, Bologna, Italy \label{inst18}
\and
INAF - Osservatorio Astronomico di Palermo, Piazza del Parlamento 1, 90134, Palermo, Italy \label{inst16}
\and
ASI Science Data Center, Via del Politecnico SNC, 00133 Roma, Italy \label{inst21}
\and
Instituto de F\'isica y Astronomi\'ia, Universidad de Valparai\'iso, Chile \label{inst25} 
\and
N\'ucleo de Astronom\'{i}a, Facultad de Ingenier\'{i}a, Universidad Diego Portales, Av. Ej\'ercito 441, Santiago, Chile \label{inst24}
\and
Astrophysics Research Institute, Liverpool John Moores University, 146 Brownlow Hill, Liverpool L3 5RF, United Kingdom \label{inst15}
\and
Max-Planck Institut f\"{u}r Astronomie, K\"{o}nigstuhl 17, 69117 Heidelberg, Germany \label{inst22}
\and
Department of Physics and Astronomy, Uppsala University, Box 516, SE-751 20 Uppsala, Sweden \label{inst23}
\and
INAF - Padova Observatory, Vicolo dell'Osservatorio 5, 35122 Padova, Italy \label{inst14}
 }

   \date{Received...; accepted 30 March 2017}

   \newcommand{\teff}{$T_{\rm eff}~$}
   \newcommand{\logg}{$\log{g}~$}
   \newcommand{\feh}{$\rm [Fe/H]~$}
   \newcommand{\met}{${\rm [M/H]}~$}
   \newcommand{\alfafe}{${\rm [\alpha/Fe]}~$}
   \newcommand{\kms}{km~s$^{-1}~$}
   \newcommand{\vrad}{${\rm V_{rad}}~$}   
   \newcommand{\teffp}{$T_{\rm eff}$}
   \newcommand{\loggp}{$\log{g}$}
   \newcommand{\fehp}{$\rm [Fe/H]$}
   \newcommand{\metp}{${\rm [M/H]}$}
   \newcommand{\alfafep}{${\rm [\alpha/Fe]}$}
   \newcommand{\kmsp}{km~s$^{-1}$}
   \newcommand{\vradp}{${\rm V_{rad}}$}

 
  \abstract
   {As observational evidence steadily accumulates, the nature of the Galactic bulge has proven to be rather complex: the  structural, kinematic, and chemical analyses often lead to contradictory conclusions. The nature of the metal-rich bulge -- and especially of the metal-poor bulge -- and their relation with other Galactic components,  still need to be firmly defined on the basis of statistically significant high-quality data samples.}
   {We used the fourth internal data release of the Gaia-ESO survey to characterize the bulge metallicity distribution function (MDF), magnesium abundance, spatial distribution, and correlation of these properties with kinematics. Moreover, the homogeneous sampling of the different Galactic populations provided by the Gaia-ESO survey allowed us to perform a comparison between the bulge, thin disk, and thick disk sequences in the [Mg/Fe] vs. [Fe/H] plane in order  to constrain the extent of their eventual chemical similarities.}
   {We obtained spectroscopic data for $\sim2500$ red clump stars in 11 bulge fields, sampling the area $-10^\circ\leq l\leq+8^\circ$ and $-10^\circ\leq b\leq-4^\circ$ from the fourth internal data release of the Gaia-ESO survey. A sample of $\sim6300$ disk stars was also selected for comparison. Spectrophotometric distances computed via isochrone fitting allowed us to define a sample of stars likely located in the bulge region.}
   {From a Gaussian mixture models (GMM) analysis, the bulge MDF is confirmed to be bimodal across the whole sampled area. The relative ratio between the two modes of the MDF changes as a function of $b$, with metal-poor stars dominating at high latitudes. The metal-rich stars exhibit bar-like kinematics and display a bimodality in their magnitude distribution, a feature which is tightly associated with the X-shape bulge. They overlap with the metal-rich end of the thin disk sequence in the [Mg/Fe] vs. [Fe/H] plane. On the other hand, metal-poor bulge stars have a more isotropic hot kinematics and do not participate in the X-shape bulge. Their Mg enhancement level and general shape in the [Mg/Fe] vs. [Fe/H] plane is comparable to that of the thick disk sequence. The position at which [Mg/Fe] starts to decrease with [Fe/H], called the ``knee'', is observed in the metal-poor bulge at $\textmd{[Fe/H]}_\textmd{knee}=-0.37\pm0.09$, being 0.06 dex higher than that of the thick disk. Although this difference is inside the error bars, it suggest a higher star formation rate (SFR) for the bulge than for the thick disk. We estimate an upper limit for this difference of $\Delta\textmd{[Fe/H]}_{knee}=0.24$ dex. Finally, we present a chemical evolution model that suitably fits the whole bulge sequence by assuming a fast ($<1$ Gyr) intense burst of stellar formation that takes place at early epochs.}
   {We associate metal-rich stars with the bar boxy/peanut bulge formed as the product of secular evolution of the early thin disk. On the other hand, the metal-poor subpopulation might be the product of an early prompt dissipative collapse dominated by massive stars. Nevertheless, our results do not allow us to firmly rule out the possibility that these stars come from the secular evolution of the early thick disk. This is the first time that an analysis of the bulge MDF and $\alpha$-abundances has been performed in a large area on the basis of a homogeneous, fully spectroscopic analysis of high-resolution, high S/N data.}

   \keywords{Galaxy: bulge, formation, abundances, stellar content -- stars: abundances
               }
   \maketitle
   
\section{Introduction}
\label{sec:introduction}

The Galactic bulge is the Rosetta stone for our understanding of galaxy formation and evolution. Being a major Galactic component, comprising around a quarter of the Milky Way stellar mass \citep[$M_\textmd{bulge}=2.0\pm0.3\times10^{10} M_\odot$,][]{valenti2016}, and covering around 500-600 square degrees in the sky, the Galactic bulge provides us with the closest example of this kind of frequent galactic structure. As a predominantly old stellar population \citep{zoccali2003,clarkson2008}\footnote{\citet{clarkson2011} studied the CMD of proper motion selected bulge stars with HST photometry, and found that only $\leq3.4$\% of the bulge population can be younger than 5 Gyr. However, from the spectroscopic analysis of a sample of lensed bulge dwarfs, \citet{bensby2013} found that nearly 22\% are younger than 5 Gyr.}, it witnessed the very early formation history of the Milky Way, and its stars contain a detailed record of the past chemodynamical events that shaped its current observable properties. This valuable information can be read from photometric and/or spectroscopic observations of its resolved stellar populations. The star-by-star study of its stellar content, together with the great degree of detail that is possible to achieve with the current large aperture telescopes and multiobject spectroscopy, have turned the bulge into an opportunity to perform near field cosmology in order to test any envisaged scenario of galaxy formation.

Currently, there are two broad scenarios of bulge formation. The first assumes an early prompt formation, whether through a dissipative collapse of a primordial cloud contracting in a free-fall time \citep{eggen1962} or through the accretion of substructures, disk clumps, or external building blocks in a $\Lambda \textmd{CDM}$ context \citep{scannapieco2003,immeli2004}. The predicted outcome of this process is a classical bulge, a centrally concentrated spheroidal structure, predominantly made up of old stars and dynamically sustained by isotropic random orbital motions. The second scenario conceives the bulge formation as the product of secular internal evolution of the early disk over longer timescales. In this case, dynamical instabilities of the early inner disk lead to the formation of a bar, a structure  which subsequently undergoes vertical instabilities, buckling, and redistributing disk angular momentum in the vertical direction. The resulting structure --  which has a characteristic boxy peanut (B/P)  or, in extreme cases, an X-shaped morphology -- is commonly called a pseudobulge.

In the last decade, the study of the Milky Way bulge has experienced a revolution, mainly driven by technical improvements in instrumentation and telescope aperture, allowing the execution of several mid- and large-scale spectroscopic and photometric surveys of the central Galactic region. The complex picture that has emerged from this very active research makes it evident that the Galactic bulge can no longer be considered a simple homogeneous structure.

The Galactic bulge hosts a bar \citep[e.g.,][]{devaucouleurs1964,liszt1980,weiland1994}, currently characterized as a triaxial structure of $\sim3.5$ kpc in length flaring up into an X-shape structure \citep{wegg2013,Ness2016}. This configuration is predicted as an outcome of secular disk evolution.

On the other hand, the metallicity distribution function (MDF) study by \citet{zoccali2008} demonstrated the existence of a vertical metallicity gradient along the bulge minor axis in the range $b=[-4:-12]^\circ$. This gradient, already suggested by \citet{minniti1995}, was interpreted as the signature of classical bulge formation. Using the same sample, \citet{babusiaux2010} showed that metal-rich stars present a vertex deviation compatible with bar driven kinematics. Instead, the metal-poor component exhibits isotropic kinematics, as expected for a classical spheroid. The work of \citet{hill2011} on Baade's window, revealed that these kinematical signatures can be correlated with a bimodal nature of the MDF, which is also found in other fields \citep{Uttenthaler2012,rojas-arriagada2014,gonzalez2015,zoccali2017}. The work of \citet{ness2013} challenged this picture from their analysis of $\sim10200$ likely bulge stars from the ARGOS survey. In fact, their MDFs from $l=\pm15^\circ$ strips at $b=-5^\circ, \,-7.5^\circ, \,-10^\circ$ are trimodal. They related the double red clump feature, a signature of the B/P bulge, only with $\textmd{[Fe/H]}\geq-0.5$ stars, and the vertical metallicity gradient with a change in the relative size of the metallicity components. The determination of the intrinsic shape of the bulge metallicity distribution function is fundamental because its exact multimodal shape can be related with a number of different bulge formation channels.

In this general context, attempts to conciliate morphological, chemical, and kinematical evidence argue for a composite nature of the bulge. Recent research seems to agree on the bar-driven secular origin of the metal-rich bulge. Secular evolution through disk instability is able to reproduce the chemical, morphological, and kinematic properties displayed by bulge stars in this metallicity range. Instead, there is less consensus on the origin of the metal-poor bulge. Its spatial distribution seems to be uncorrelated with the bar position, appearing as an extended, centrally concentrated and possibly spheroidal component. This is supported by the distribution found for other tracers of metal-poor old populations such as RR Lyrae stars (\citeauthor{pietrukowicz2012} \citeyear{pietrukowicz2012}, \citeauthor{dekany2013} \citeyear{dekany2013}, \citeauthor{Kunder2016} \citeyear{Kunder2016}, \citeauthor{Gran2016} \citeyear{Gran2016}, but see also \citeauthor{pietrukowicz2015} \citeyear{pietrukowicz2015}). On the chemical abundance side, $\alpha$-abundance ratios with respect to iron are systematically enhanced over its whole metallicity range.

 \begin{figure*}[]
 \centering
 \includegraphics[width=13.5cm]{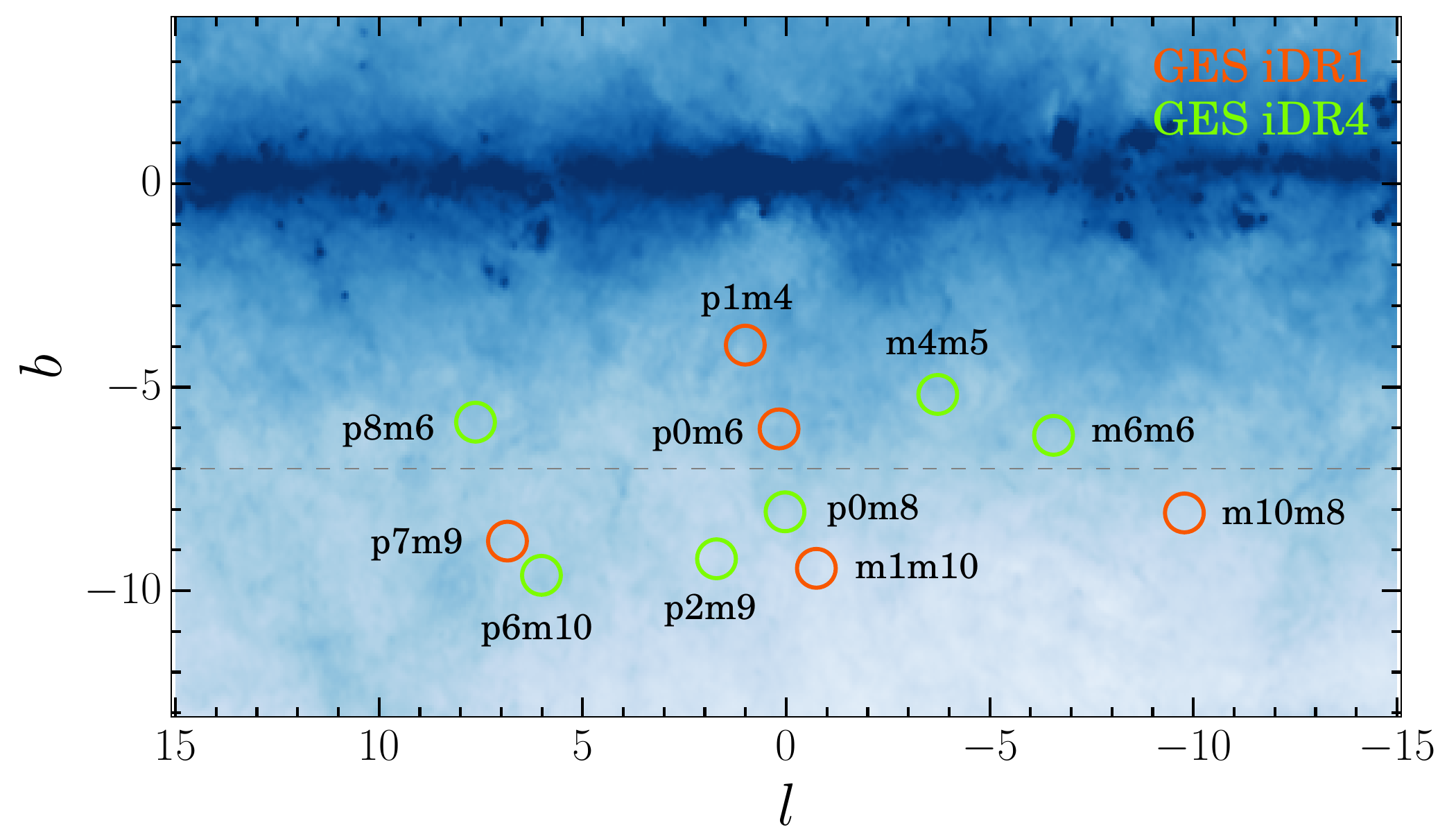}
 \caption{Position of the 11 bulge fields analyzed in the present study. The five red circles indicate the fields already examined in \citet{rojas-arriagada2014} (Gaia-ESO survey iDR1), while the six green circles show  the extra fields observed up to the iDR4. Each field is labeled according to the name coding adopted throughout the paper and based on the Galactic coordinates. The background image corresponds to an extinction map of the bulge region according to the \citet{schlegelMapas} prescription. The blue color density code saturates close to the plane where the extinction is high. A horizontal dashed gray line indicates $b=-7^\circ$, used to divide the sample into fields close to and far from the plane.} 
 \label{fig:fields_distribution}
 \end{figure*}

Detailed comparisons between bulge and thick disk samples in the  [$\alpha$/Fe] vs. [Fe/H] plane provide a direct way to try to understand the origin of the metal-poor bulge. Early attempts in this direction \citep{zoccali2006,lecureur2007,fulbright2007} claimed that the bulge presents higher $\alpha$-enhancements relative to the thick disk. \citet{melendez2008} and \citet{alves-brito2010} attributed this result to systematic effects arising from the comparison of giant and dwarf samples given their different temperature and gravity regimes. Their homogeneous sample of bulge and local thick disk giants display chemical similarities, with similar trends in the [$\alpha$/Fe] vs. [Fe/H] plane, and presumably a comparable location of the  so-called ``knee'' in the sequences of both populations. Similarities between the bulge and the thick disk have also  been  suggested using dwarf stars \citep{bensby2013, bensby2014}. The study of the detailed chemical abundance patterns from statistically significant homogeneously analyzed samples can shed light on the initial conditions, physical processes, and relative timescales characterizing  formation and evolution of the bulge and thick disk populations.

All in all, the puzzle of bulge formation has many pieces, and not all of them are currently in their definitive place. In this paper, we   provide new evidence on  some of the issues discussed above. To this end, we made use of data coming from the fourth internal data release of the Gaia-ESO survey (iDR4). The Gaia-ESO survey is a large ongoing public spectroscopic survey (300 nights from the end of 2011 to the end of 2016) targeting $\sim10^5$ stars distributed in all the main components of the Milky Way: the halo, bulge, and the disk system \citep{GESMessenger}. The present study is an extension of our previous work  \citep{rojas-arriagada2014}, which was based on a subset of the fields studied here and not including the analysis of individual abundances. The structure of the paper is as follows. In Sect.~\ref{sec:datos} the data are presented, the selection function of the Gaia ESO survey described, and the data processing outlined. In Sect.~\ref{sec:distance_reddening} we present the method and the results obtained for stellar distances and reddening determinations from an isochrone fitting procedure. The bulge metallicity distribution function is presented in Sect.~\ref{sec:MDFs}, while the trends in the [Mg/Fe] vs. [Fe/H] and correlations with kinematics in Sect.~\ref{sec:bulge_trends}. A search for chemical similarities between the bulge and the thick disk is presented in Sect.~\ref{sec:chem_simil_bulge_thick_disk}. A comparison with a chemical evolution model is presented in Sect.~\ref{sec:comparison_CEM}. Finally, the discussion and our conclusions are drawn in Sect.~\ref{sec:discussion}.

\section{Data}
\label{sec:datos}

In the present study, we made use of data coming from the fourth internal data release of the Gaia-ESO survey. The Gaia-ESO survey consortium is based on working groups in charge of the different tasks, from target selection and observation to the derivation of the different fundamental parameters and abundances required to achieve the scientific goals of the survey. A general description of the survey can be found in \citet{GESMessenger}, while a description of the data processing flow is briefly outlined below. 

We work with a sample of 2320 red clump stars from observations collected up to the iDR4 of the Gaia-ESO survey. They are distributed in 11 pointings toward the bulge region. The positions of the observed fields\footnote{We refer to the fields throughout the paper by a name convention using their Galactic longitude $l$ and latitude $b$, and the p/m letter coding the $\pm$ sign, to assemble their names. For example, the field at $l=7$ $b=-9$ is named p7m9.} are illustrated in Fig.~\ref{fig:fields_distribution} overplotted on top of an extinction map of the bulge region \citep[using data from the extinction maps of][]{schlegelMapas}. Five of the fields were already observed during the first nine months of the Gaia-ESO survey project, and released in the iDR1. They were analyzed in a previous Gaia-ESO survey publication \citep{rojas-arriagada2014}, although the $\alpha$-abundances were not included in iDR1. For comparison purposes, a sample of 228 red giant branch (RGB) and red clump stars in Baade's window was adopted from \citet{zoccali2008} and \citet{hill2011}. These comparison stars were reobserved and analyzed in the same way as the rest of the Gaia-ESO survey bulge targets, and added to the main sample making a total of 2548 stars. In addition, a set of spectroscopic benchmark stars were observed to calibrate the computed spectroscopic parameters. Spectra were obtained with the ESO/VLT/FLAMES facility \citep{pasquini2000} in the MEDUSA mode of the GIRAFFE multi-object spectrograph. Only the HR21 setup was employed (except for half of the stars in the comparison sample, which were also observed with the HR10 setup), providing a spectral coverage spanning from 8484 to 9001 \AA\ with a resolving power of  $R\sim16200$. The general quality of the obtained spectra is quite good: the  average signal-to-noise ratio (S/N) is 290 and no spectrum has less than 80 per resolution element.

 \begin{figure}
 \begin{center}
 \includegraphics[width=8.7cm]{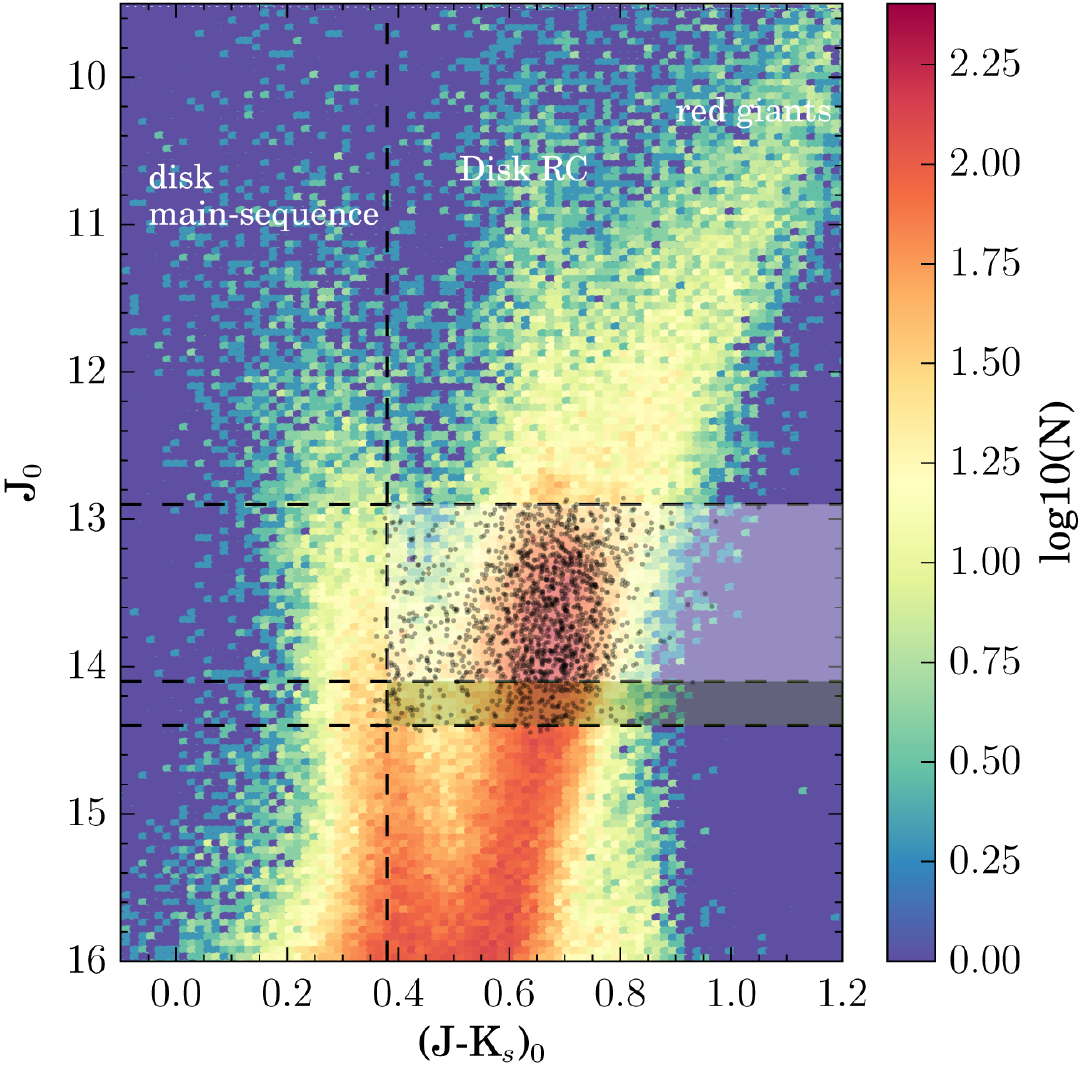}
 \caption{Gaia-ESO Survey bulge selection function. The background Hess diagram depicts a generic CMD in the bulge region from VVV photometry. Prominent sequences are labeled. A shaded white area indicates the main selection function, with  color and magnitude cuts of $(\textmd{J-K}_s)_0>0.38$ and $12.9<\textmd{J}_0<14.1$, respectively. The green shaded area indicates the magnitude extension implemented in fields where the double red clump feature is visible. The whole spectroscopic sample analyzed in the present study is displayed with black dots.} 
 \label{fig:photom_selection}
 \end{center}
 \end{figure}

\subsection{Target selection}
\label{subsec:target_sel}
The targets were selected with a photometric selection function specifically designed for the bulge portion of the Gaia-ESO survey. It made use of $J$ and $K_s$ photometry available from the Vista Variables in the Via Lactea project \citep[VVV;][]{minniti2010}. This selection is illustrated in Fig.~\ref{fig:photom_selection}. A generic color cut selects stars with $(\textmd{J-K}_s)_0>0.38$ mag, which is imposed on the dereddened photometry in each field according to the values estimated from the reddening map of \citet{gonzalez2011b}\footnote{These maps, derived from VVV and 2MASS photometric data, are accessible at \url{http://mill.astro.puc.cl/BEAM/calculator.php}.}. This cut, defining the left border of the selection box in Fig.~\ref{fig:photom_selection}, is blue enough to allow metal-poor bulge stars to be included in the sample, but has  the drawback of including a number of foreground dwarf main-sequence stars. This sample contamination enters in a variable proportion according to the field extinction. The latter because the dwarf disk stars are distributed in the CMD mostly in a vertical band at the blue side of the bulge RC. This blue plume is on average less affected by the reddening than the bulge RC, so that the difference in color between the two features depends on the specific field extinction. In Fig.~\ref{fig:photom_selection}, the dwarf thin disk plume is visible at $J-K_0\sim0.35$ mag, while that corresponding to the disk RC at $J-K_0\sim0.65$ mag. Since RC stars are good standard candles, the RC sequence clumps in magnitude whenever these stars clump spatially. This happens at $J_0=13.5$ mag which, in fact, corresponds to the mean apparent magnitude of a RC star located in the Galactic bulge.
On the other hand, a generic magnitude cut selects stars with $(12.9< \textmd{J}_0<14.1$ mag. This 1.2 mag interval is in general large enough to select stars located in the bulge RC peak of the field luminosity function, accounting for the spatial distance spread of the bar and the change in mean magnitude with longitude because of the bar position angle. In a number of fields where a double RC is observed in the luminosity function, the magnitude cut would not fit the entire magnitude extension of the bar. In these cases, an extension of the magnitude limit was allowed to include up to 30\% of the targets in an extra 0.3 mag below the nominal cut. 

  \begin{figure}
 \begin{center}
 \includegraphics[width=8.6cm]{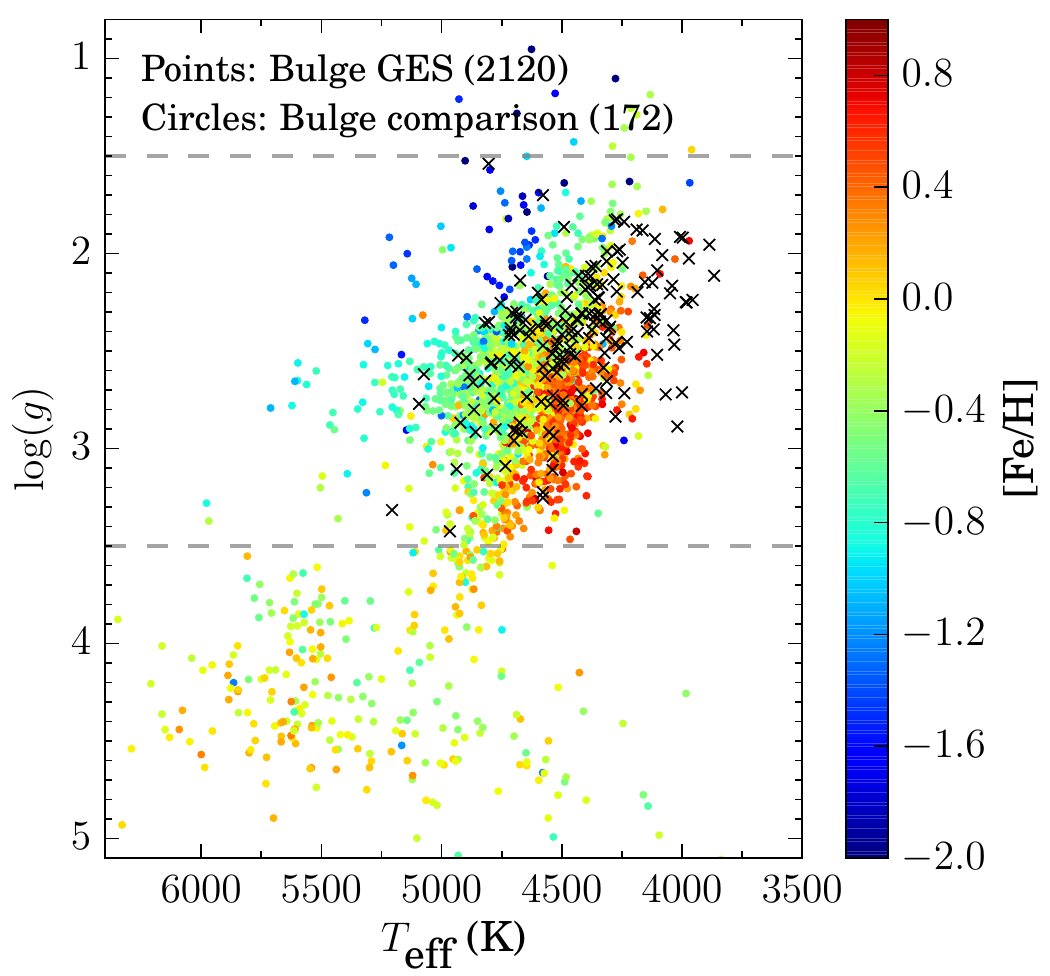}
 \caption{HR diagram of the bulge sample stars for which iron determinations from FeI lines are available (2292 out of 2548 stars). Stars selected with the Gaia-ESO photometric selection function are indicated as full circles color-coded by metallicity. The subset of RC and RGB comparison stars are indicated by black crosses. Two dashed gray lines mark \logg=1.5 and 3.5 dex.} 
 \label{fig:HR_bulge}
 \end{center}
 \end{figure}

The above selection function draws the main sample of 2320 RC stars. Instead, the sample of 228 comparison stars have selection functions described in \citet{zoccali2008} and \citet{hill2011}.

\begin{figure}
 \begin{center}
 \includegraphics[width=7.5cm]{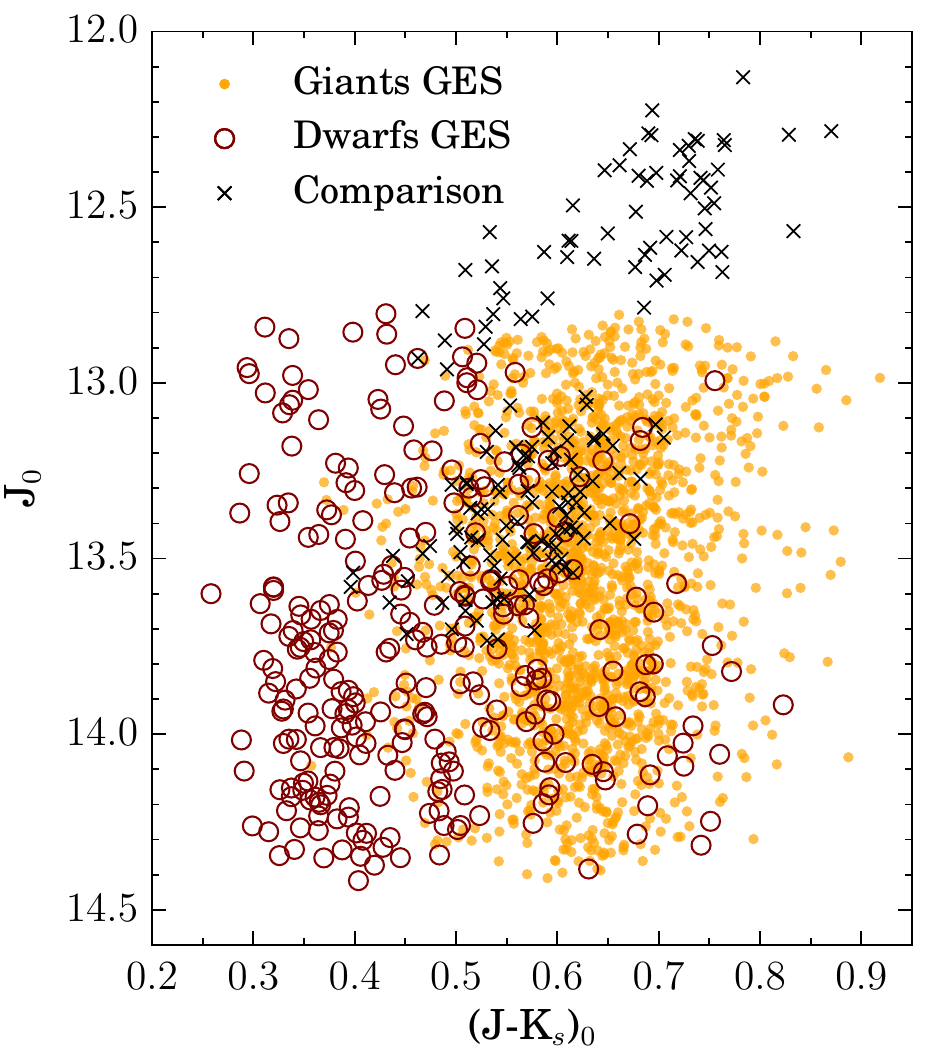}
 \caption{Color magnitude diagram of the whole spectroscopic sample. Stars with $\log(g)>3.5$ are marked as open brown  circles, while those with $\log(g)<3.5$ as filled orange circles. The comparison sample of RC and RGB stars (all with $\log(g)<3.5$) are indicated by black crosses.} 
 \label{fig:CMD_ginats_dwarfs}
 \end{center}
 \end{figure}

\subsection{Radial velocities, stellar parameters, and individual abundances}
\label{subsec:rv_params_and_abund}
Radial velocities are measured by the Gaia-ESO survey with a dedicated pipeline by cross-correlation against real and synthetic spectra (Koposov et al., in prep). In our sample, velocity uncertainties are lower than $0.4$~\kms.
The determination and compilation of a recommended set of atmospheric parameters and elemental abundances is performed by the Gaia-ESO survey working group 10 (WG10) for all the F-, G-, and K-type stars observed with GIRAFFE. A detailed description of the process will be published in Recio-Blanco et al. (in prep). In short, the individual spectra are analyzed  using three independent approaches: Spectroscopy Made Easy \citep[SME;][]{valenti1996}, FERRE \citep{allende-prieto2006}, and MATISSE \citep{recio-blanco2006}. This is performed in a model-driven way by comparing the observed spectra against synthetic templates, whether interpolated from a dense grid or computed on the fly. In this way \teff, \logg, [M/H], and [$\alpha$/Fe] are determined by the three nodes. A set of spectroscopic benchmark stars \citep{jofre2015} is analyzed in the same way. For each node, the differences between the calculated and the nominal fundamental parameters are estimated for the set of benchmark stars. Using these values, the node results for a given program star are bias corrected into the astrophysical scale given by the benchmark stars, and then combined in average to produce a unique set of atmospheric parameters while reducing the random errors of individual determinations. The corresponding  errors are computed as the node-to-node dispersion in order to properly account for large node-to-node discrepancy in  low-quality parametrization. They constitute the recommended set of model-driven, multi-method fundamental parameters by the Gaia-ESO survey consortium.

\begin{table} 
\centering
\small
\caption{Characterization of the observed fields. Signal-to-noise ratios are the field average. E(J-K)$_{G11}$ corresponds to the reddening as computed from the extinction maps of \citet{gonzalez2011b} in a box of 30 arcmin per side centered in the respective ($l$, $b$) coordinates. E(J-K)$_{fit}$ values are those estimated in Sect.~\ref{sec:distance_reddening} from isochrone fitting. Finally, N$_{\textmd{G}}$/N$_{\textmd{fld}}$ provides the ratio between giant ($\log(g)<3.5$) and the total number of stars per field.}
\begin{tabular}{lcccccc}
\hline
\hline
Field  & $l$ & $b$ & $\overline{\textmd{SN}}$ & E(J-K)$_{G11}$ & E(J-K)$_{fit}$ & N$_{\textmd{G}}$/N$_{\textmd{fld}}$  \\
name       &     &     &                          &                &                 &                                    \\\hline
p1m4  &  1.00 & -3.97 & 364 & 0.26 & 0.20 & 359/369 \\
p0m6  &  0.18 & -6.03 & 254 & 0.14 & 0.17 & 180/204 \\
m1m10 & -0.74 & -9.45 & 340 & 0.03 & 0.06 & 131/187 \\
p7m9  &  6.85 & -8.87 & 244 & 0.10 & 0.11 & 200/221 \\
m10m8 & -9.78 & -8.09 & 347 & 0.03 & 0.08 & 234/310 \\
m4m5  & -3.72 & -5.18 & 192 & 0.19 & 0.18 & 89/94   \\
m6m6  & -6.57 & -6.18 & 168 & 0.13 & 0.13 & 189/206 \\
p0m8  &  0.03 & -8.06 & 310 & 0.06 & 0.07 & 81/98   \\
p2m9  &  1.71 & -9.22 & 362 & 0.08 & 0.08 & 81/105  \\
p8m6  &  7.63 & -5.86 & 250 & 0.22 & 0.20 & 284/302 \\
p6m9  &  6.01 & -9.62 & 279 & 0.09 & 0.08 & 159/196 \\\hline
\end{tabular}
\label{tab:caract_campos}
\end{table}

This set of parameters, the Gaia-ESO survey linelist  used to compute the synthetic spectra in the previous step \citep{heiter2015}, and the MARCS model atmospheres \citep{gustafsson2008}, are adopted to determine the elemental abundances of $ \alpha$- and iron-peak elements (including the iron and magnesium abundances used in this work)  using SME and an automated spectral synthesis method \citep{mikolaitis2014}. The results from the two methods compare well, and only small bias corrections are needed. The final abundances for each element are calculated as the average of the two individual determinations, while errors are taken proportional to the absolute difference between them. Finally, abundances relative to the Sun are derived by adopting the solar composition of \citet{grevesse2007}. They constitute the recommended set of abundances by the Gaia-ESO survey consortium.

It is worth  highlighting here that, contrary to the Gaia-ESO survey iDR1 \citep[used in our previous bulge study][]{rojas-arriagada2014}, the procedure described above includes three improvements: (1) the use of three codes instead of one to compute the fundamental parameters, thus providing final results with smaller statistical, and hopefully, systematic errors; (2) the availability of elemental abundances which enable us to perform a more detailed analysis than that presented in \citet{rojas-arriagada2014}; and (3) a more robust calibration of both the stellar parameters and abundances thanks to a larger sample of observed benchmarks.

Although the present sample contains a number of fields already studied from the iDR1, the fundamental parameters and abundances adopted here come from the iDR4, as is true for the rest of the sample.
 
In Fig.~\ref{fig:HR_bulge}, we display the Hertzsprung-Russell (HR) diagram, using the fundamental parameters of bulge stars for which the iron determinations from FeI lines are available (almost all of which also have  Mg measurements). We can verify the general good quality of the stellar parametrization because  the main HR features, main sequence, turn-off and red clump are clearly distinguishable. It is also apparent that the nature of the Gaia-ESO survey selection function leads, as anticipated, to a sample with some contamination from dwarf main-sequence stars. We select stars with $\log(g)<3.5$ (and $\log(g)>1.5$ to avoid giants for which stellar parametrization could suffer from modeling uncertainties) as our RC bulge sample. The dichotomy between the RC and dwarf stars is explicitly shown in the CMD diagram in Fig.~\ref{fig:CMD_ginats_dwarfs}. The figure clearly shows that  the dwarf contaminants are preferentially located on the blue side of the CMD toward the locus where the blue plume of disk dwarf stars is visible in a general bulge field CMD (cf. Fig.~\ref{fig:photom_selection}). A small number of dwarf stars are visible at (J-K$_s$)$_0\gtrsim0.65$. They correspond to a fraction of the stars with $\log(g)$ values that are slightly higher than the cut at $\log(g)=3.5$ dex. Although they could correspond to RC members according to their colors, we adopt their spectroscopic classification. This good general correspondence between stars in the HR and CMD diagrams constitutes a sanity check on the internal consistency of the stellar parametrization.

In the following we made use of the giant/RC sample defined above. It contains mostly RC with a contribution of RGB stars, and is composed of 1987 stars (including stars from the comparison sample).

\section{Distance and reddening estimations}
\label{sec:distance_reddening}

We calculated individual line-of-sight  spectrophotometric distances and reddenings for the whole sample with available FeI measurements (2273 stars). The adopted procedure made use of the fundamental parameters \teff; \logg; [Fe/H] (from FeI lines); and VISTA $J$, $H$, and $K_s$ photometry and associated errors to compute simultaneously the most likely line-of-sight distance and reddening  by isochrone fitting with a set of PARSEC isochrones\footnote{Available at \url{http://stev.oapd.inaf.it/cgi-bin/cmd}}. The general approach, rather similar to other methods in the literature \citep[e.g.,][]{zwitter2010,Ruchti2011,Kordopatis2011}, is outlined below.
 
 \begin{figure}
 \begin{center}
 \includegraphics[width=8.4cm]{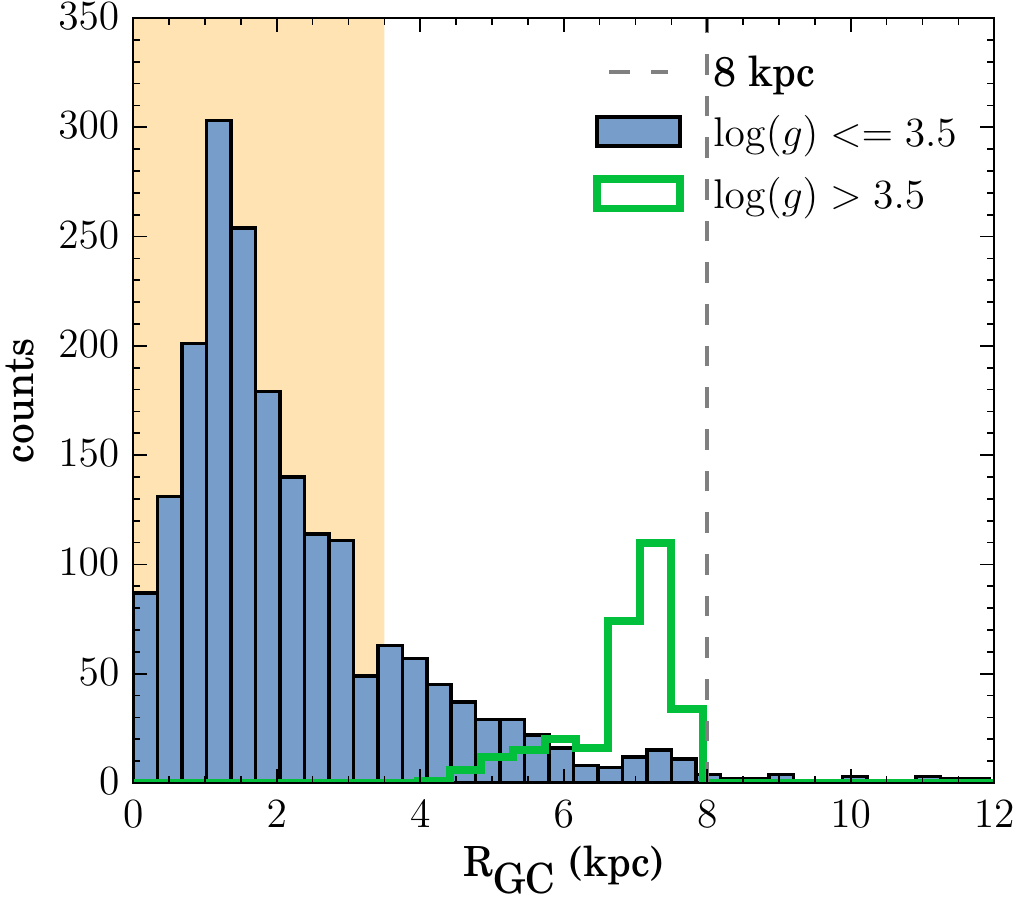}
 \caption{Distribution of Galactocentric radial distances of RC (blue bars) and dwarf (green profile) stars. A shaded yellow area highlight the spatial cut ($R_{GC}<3.5$ kpc) adopted to define our bulge working sample. A vertical dashed gray line indicates the solar Galactocentric radius.}
 \label{fig:histo_dist_bulge}
 \end{center}
 \end{figure}

\begin{itemize}
 \item[1.] We consider a set of isochrones spanning ages from 1 to 13~Gyr in steps of  1~Gyr  and metallicities from $-2.2$ to $+0.5$ dex in steps of 0.1 dex. In practice,  for a given age and metallicity, each isochrone consists of a sequence of model stars with increasing mass located along a track in the \teff vs. \logg plane from the main-sequence to the AGB. Each model star is characterized by theoretical values of the absolute magnitudes $M_J$, $M_H$, and $M_{K_s}$. On the other hand, an observed star is characterized by a vector containing a set of fundamental parameters and observed passband magnitudes $\{T_\textmd{eff},\log(g), \textmd{[Fe/H]},J, H, K_s\}$, together with their associated errors. Given the three fundamental parameters \teff, \logg, and [Fe/H], a star can be placed in the isochrone \teff-\logg-[Fe/H] space. 
 
 \item[2.] We compute the distance from this observed star to the whole set of model stars considering all the isochrones. To this end, we adopt the metric
\begin{align*}
d(a.m)&=\frac{\left[ T_{\textmd{eff}\ *}-T_\textmd{eff}(a,m)\right]^2}{\sigma_{T_\textmd{eff}\ *}^2} + \frac{\left[ \log(g)_{*}-\log(g)(a,m)\right]^2}{\sigma_{\log(g)\ *}^2} \\ & \hspace{11pt}+\frac{\left[ \textmd{[Fe/H]}_{*}-\textmd{[Fe/H]}(a,m)\right]^2}{\sigma_{\textmd{[Fe/H]}\ *}^2},
\end{align*}
where $T_\textmd{eff}(a,m)$, $\log(g)(a,m)$, and $\textmd{[Fe/H]}(a,m)$ are the fundamental parameters, depending on the age $a$ and mass $m$, characterizing the isochrone model stars. The quantities with a star subscript stand for the fundamental parameters and errors ($\sigma_{T_\textmd{eff}\ *},\ \sigma_{\log(g)\ *}$) of the observed star. 

\item[3.] Using this metric, we compute weights associated with the match of the observed star with each point of the isochrone collection
$$W(a,m)=P_m P_{\tiny{IMF}}\left[e^{-d(a,m)}\right].$$
This weight is composed of three factors:
\begin{itemize}
 \item[a.] $P_m$ accounts for the evolutionary speed of the model stars along the isochrone. The isochrones are constructed in order to roughly distribute their model stars uniformly  along them. This means that a simple unweighted statistic using all the model stars will lead to overweight short evolutionary stages and not long-lived ones. A way to correct for this effect is to include a weight $P_m$ proportional to the $\Delta m$ between contiguous model stars in order to assign more weight to the long-lived evolutionary stages where a randomly selected star is more likely to be;
 \item[b.] $P_{\tiny{IMF}}$ accounts for the fact that, given a stellar population, the number of stars per mass interval $dN/dm$ is not uniform. In fact, this distribution is given by the initial mass function (IMF)\footnote{In practice, we made use of the PARSEC isochrone quantity \textit{int\_IMF}, which is the cumulative integral of the IMF along the isochrone. In fact, following \citet{girardi2000}, we assume that ``the difference between any two values is proportional to the number of stars located in the corresponding mass interval''.}; 
 \item[c.] The third factor is  an exponential weight associated with the distance of the observed star with respect to each model star, given the adopted metric. We can use the weights $W(a,m)$ to compute any kind of weighted statistics.
 \end{itemize}
 
 \item[4.] We calculate for a given observed star the likely values of its absolute magnitudes $M_J$, $M_H$, and $M_K$ from the set of isochrones.
  
 \item[5.] We compute the line-of-sight reddening by comparing the theoretical color with the observed color $E(J-K)=(J_{obs}-K_{obs})-(M_J-M_K)$.
 
  \item[6.] Finally, from these values, by considering the observed photometry $J$, $H$, $K_s$ and the estimated reddening, we compute the distance modulus and then line-of-sight distances.
\end{itemize}

We  computed distances and reddening values (field averages are quoted in  Col.~6 in Table~\ref{tab:caract_campos}) for the whole bulge sample with available [Fe/H] values. Typical internal errors in distance are  about 25-30\%. Using the ($l$,$b$) star positions, we also compute the Galactocentric Cartesian coordinates $X_{GC}$, $Y_{GC}$, and $Z_{GC}$, and the cylindrical Galactocentric radial distance $R_{GC}=\sqrt{X_{GC}^2+Y_{GC}^2}$. The distribution of the latter is shown in Fig.~\ref{fig:histo_dist_bulge}, separately for the giant and dwarf portions of our sample. We can see how the stars we found to be foreground dwarf contaminants based on their \logg values are in fact located mainly at 7-8 kpc, in the solar neighborhood. On the other hand, the presumed RC bulge stars are found in a narrow distribution with a peak at $\sim1.5$~kpc. We did not expect this maximum to be at $R_{GC}=0$~kpc given that most of our fields are several degrees apart from the Galactic plane. The shape of the $R_{GC}$ distribution led us to introduce a radial distance cut, defining a working sample of likely bulge stars. To this end, we adopted the criterion $R_{GC}\leq3.5$~pc. This restriction was applied to the giant sample already defined from their \logg values. The resulting working sample is composed of 1583 stars.

\section{Metallicity distribution function}
\label{sec:MDFs}

 \begin{figure}[!ht]
 \begin{center}
 \includegraphics[width=8.6cm]{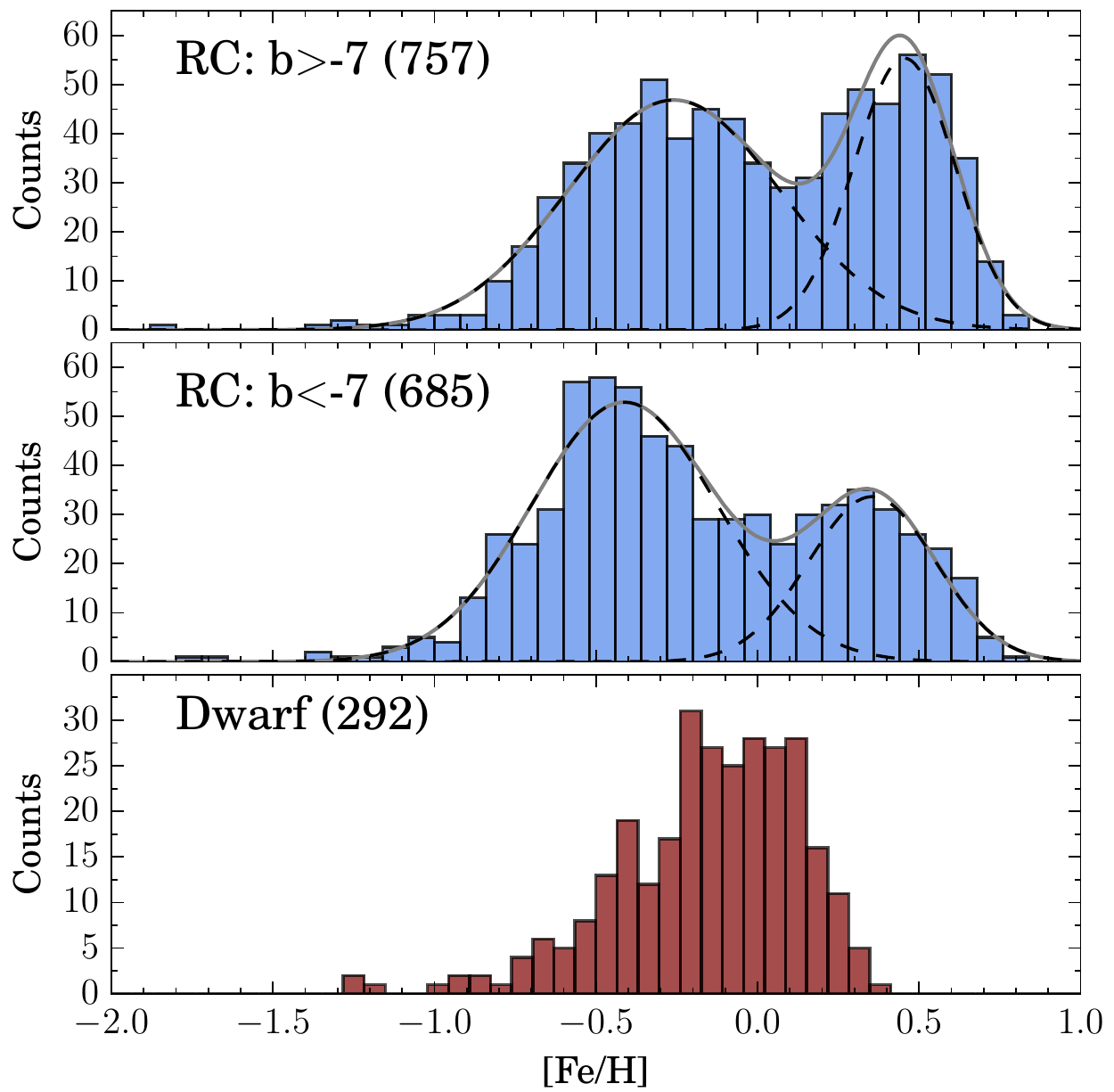}
 \caption{\textit{Upper panel:} Combined MDF of fields located close to the Galactic plane ($b>-7^\circ$). \textit{Middle panel:} Combined MDF of fields located far from the Galactic plane ($b<-7^\circ$). The individual GMM components are drawn with black dashed lines, while their combined profile as a solid gray line. \textit{Lower panel:} MDF of stars classified as dwarfs according to their \logg values. In all panels, the total number of stars is indicated in parentheses.} 
 \label{fig:MDF_bulge_disk}
 \end{center}
 \end{figure}

We studied the shape of the MDF from our working sample of likely bulge stars, excluding the comparison stars because their different selection function might bias the MDF toward high metallicity. As a first glimpse of the bulge MDF, we split the sample into two groups of fields which are close to or far from the plane. They are a combination of fields located at $b>-7^\circ$ and $b<-7^\circ$, respectively (the horizontal dashed gray line in Fig.~\ref{fig:fields_distribution}). In this way, each  half contains a similar number of fields. While it is true that this exercise can blur specific MDF field-to-field variations, it allowed us to increase the number statistics to investigate the general characteristics of the bulge MDF. The two subsamples are displayed in the upper and middle panels of Fig.~\ref{fig:MDF_bulge_disk}. Two things are immediately apparent. First, the MDFs present a clear bimodal distribution with a narrow metal-rich component peaking at super-solar metallicities and  another broader and metal-poor component peaking at $\textmd{[Fe/H]}\approx-0.4/-0.5$~dex (in agreement with \citeauthor{hill2011} \citeyear{hill2011} and \citeauthor{gonzalez2015} \citeyear{gonzalez2015}, but in contrast with the trimodal MDF of \citeauthor{ness2013} \citeyear{ness2013}). Second, the relative proportion of stars comprising the two peaks changes with Galactic latitude. In fact, the size of the metal-rich component decreases with respect to the metal-poor one while going far from the Galactic plane. Broadly speaking, our metal-poor and metal-rich MDF components encompass the metallicity ranges ${\rm-1.0\leq[Fe/H]\leq0.0}$~dex and ${\rm0.0\leq[Fe/H]\leq1.0}$~dex. The incidence of stars with ${\rm[Fe/H]<-1.0}$~dex is low ($1.7\%$ of our sample), and given its small number we do not attempt here a detailed analysis of its properties. Accounting for our distance cut to select likely bulge members, these stars might be a combination of halo passing-by stars and the metal-poor tail of the endemic bulge population.

 \begin{figure*}
 \begin{center}
 \includegraphics[width=15.2cm]{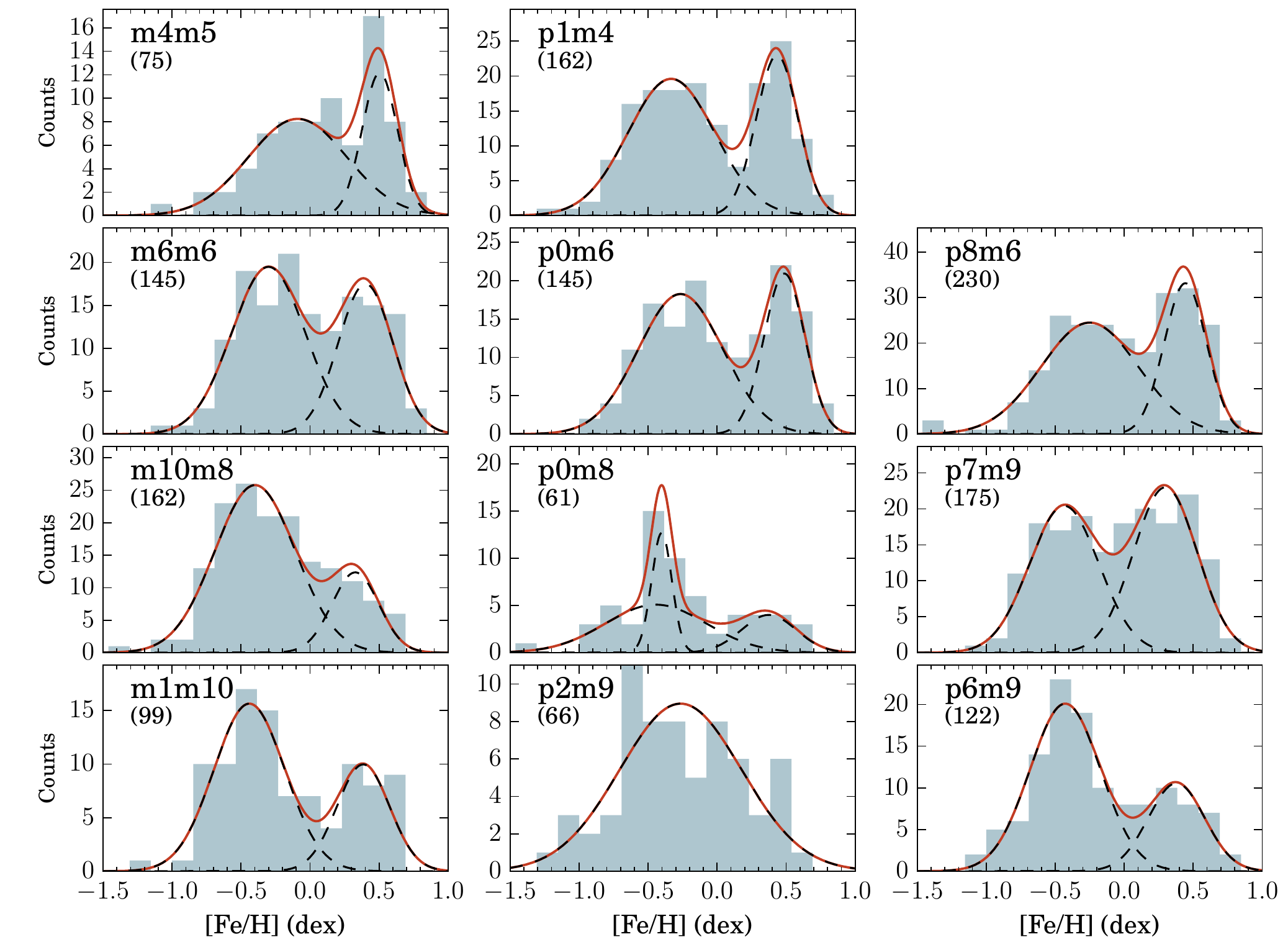}
 \caption{Metallicity distribution functions of the 11 bulge fields. Blue filled histograms stand for the individual distributions;  the number of stars is given in parentheses. An independent GMM decomposition in each field is indicated by black dashed lines (individual modes) and a red solid line (composite profile). The distribution of the fields in the panels   approximately indicates their positions in ($l$,$b$) (cf. Fig.~\ref{fig:fields_distribution}).} 
 \label{fig:MDF_fields}
 \end{center}
 \end{figure*}

To quantify these facts, as we did in \citet{rojas-arriagada2014}, we performed a Gaussian mixture models (GMM) decomposition\footnote{See \citet{ness2013} and \citet{rojas-arriagada2016} for a mathematical description of the procedure and its application to the analysis of chemical distributions of stellar populations.} on the two MDFs. In both cases, the Akaike information criterion, used for model selection, gave preference to a two-component solution with a high relative probability. Close to the plane, the narrow metal-rich component ($\sigma=0.16$~dex)   encompasses  36\% of the probability density of the model, while the broader metal-poor component  ($\sigma=0.33$~dex)  the remaining 64\%. On the other hand, far from the plane, the metal-rich ($\sigma=0.35$~dex) and metal-poor ($\sigma=0.29$~dex) components account for  30\% and  70\% of the relative weights, respectively.

As a qualitative comparison, in the lower panel of Fig.~\ref{fig:MDF_bulge_disk} we display the MDF of the sources classified as dwarfs according to their \logg values, which are mostly solar neighborhood members (Fig.~\ref{fig:histo_dist_bulge}). Their distribution resembles what it is observed in the solar neighborhood, for example by the Geneva-Copenhagen survey \citep[e.g.,][]{casagrande2011}. It is clear that these stars have a MDF with a significantly different shape with respect to the bulge sample. Their MDF has a long tail toward low metallicity (partially due to the contribution of the local thick disk) and a sharp decline toward $\textmd{[Fe/H]}=0.4$~dex. The distribution presents a strong peak at solar metallicity, precisely at the locus where the dip in the bimodality of the bulge MDF is located.

The individual MDFs of the 11 bulge fields analyzed in this work are shown in Fig.~\ref{fig:MDF_fields}. Individual GMM decompositions were attempted in each field (parameters of the best GMM fits in Table~\ref{tab:gmm_parameters}). In agreement with Fig.~\ref{fig:MDF_bulge_disk}, the preferred GMM model has two components, except in two fields (p0m8 and p2m9) where the lower number of stars prevents the GMM from giving strong statistic assessments. When comparing MDF decompositions between strips that are  at a similar latitude (rows in Fig.~\ref{fig:MDF_fields}), a decline in the number of metal-rich stars in favor of metal-poor stars with increasing distance from the Galactic plane is visible \citep[as seen also in][]{zoccali2008,ness2013}. On the other hand, while comparing fields at similar latitude, those located at positive longitudes tend to have a more enhanced metal-rich component. This asymmetry with respect to the minor axis was already characterized in the photometric metallicity map of \citet{Gonzalez2013}. As described there, it is just a perspective effect due to the bar position angle; at positive longitudes the line of sight intersects the bar at shorter distance from the plane than at negative longitudes. This means that at positive longitude our lines of sight sample regions with higher dominance of metal-rich stars than at the symmetric fields at negative longitude; consequently, the relative size of the metal-rich peak is higher, as observed in Fig.~\ref{fig:MDF_fields}.

\subsection{Quantification of metallicity gradients}
\label{subsec:quantification_met_grads}
From the GMM profiles, we first determined the metallicity at which the peaks of the two populations are located in each field (with the exception of p0m8 and p2m9). Then we computed metallicity gradients with $l$ and $b$ independently for the two populations. We also computed mean field metallicity gradients with $l$ and $b$.  For both metal-rich and metal-poor populations, we found negligible gradients with $l$ but noticeable variations with $b$  (gradients of $-0.18\ \textmd{dex/kpc}$ and $-0.31\ \textmd{dex/kpc}$, respectively). A gradient of $-0.24\ \textmd{dex/kpc}$ was found for the variation of the mean field metallicity with $b$. These values were computed by assuming all the fields centers projected on a plane at 8 kpc (to be consistent with other studies and to allow comparison). Our results are compatible with the presence of internal vertical gradients in both metallicity populations, with the gradient of the metal-poor fraction being $\sim60$ percent higher than that displayed by the metal-rich stars. In this sense, the global metallicity gradients, traditionally measured from the mean field metallicity variations with $b$, can be interpreted as the interplay of two effects: the variation of the relative proportion in which both populations contribute to the global field MDF plus the presence of internal gradients in both components. As a reference, if we compute the vertical gradient in similar fashion, but using the results for fields at $b=-4^\circ,-6^\circ,-12^\circ$ from \citet{zoccali2008}, we find a gradient of $-0.24\ \textmd{dex/kpc}$, in excellent agreement with the value we derived from our fields. Also, the photometric metallicity map of \citet{Gonzalez2013} indicates a vertical gradient of $-0.28\ \textmd{dex/kpc}$, again in agreement with the global gradient reported here.

\subsection{Spatial distribution of the subcomponents}
\label{subsec:spatial_distro_subcomponents}
\begin{figure}
 \begin{center}
 \includegraphics[width=8.5cm]{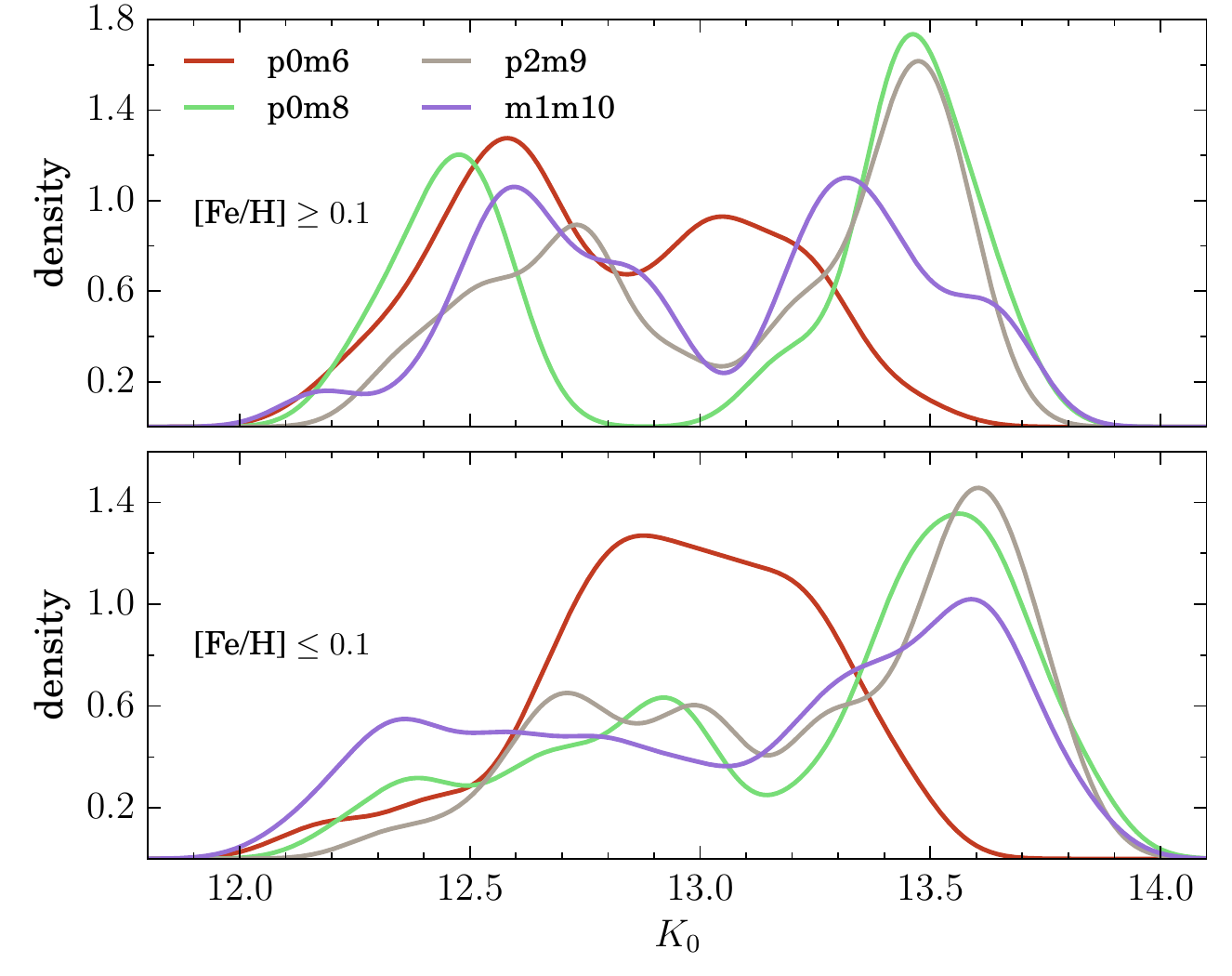}
 \caption{Double RC in the magnitude distribution of bulge stars as a function of metallicity. \textit{Upper panel: }Generalized histograms (Gaussian kernel of 0.09 mag) of the extinction corrected $K$ magnitude for stars with $\textmd{[Fe/H]}\gtrsim+0.1$~dex. \textit{Lower panel: }Generalized histograms (Gaussian kernel of 0.09 mag) of the extinction corrected $K_s$ magnitudes for stars with $\textmd{[Fe/H]}\lesssim+0.1$ dex. The same color-coding is used to identify the different fields in both panels.} 
 \label{fig:double_RC_kmag}
 \end{center}
\end{figure}

In Fig.~\ref{fig:double_RC_kmag} we display the generalized histograms of the VVV $K_s$ (reddening corrected) magnitude distributions of fields where the double RC feature is present according to the density maps of \citet{wegg2013}. The upper and lower panels show the magnitude distributions of metal-rich and metal-poor stars in each field. From the comparison of the two  sets of profiles, it is clear that an enhanced bimodality is drawn by the metal-rich stars. The difference in magnitude between the two peaks changes from field to field, being smaller closer to the plane, thus tracing the distance between the near and far arms of the X-shape bulge. On the other hand, metal-poor stars present nearly flat magnitude distributions, with some tendency, especially in the outermost fields, to have a peak at faint magnitudes. This occurs because the volume observed is bigger at greater distances, due to the cone effect.

\begin{table} 
\centering
\caption{Line-of-sight Galactocentric radial velocities of stars located in the bright and faint peaks of the metal-rich magnitude distribution. Units are in \kms.}
\begin{tabular}{llcc}
\hline
\hline
  &   & V$_{\textmd{GC}}$ bright  & V$_{\textmd{GC}}$ faint \\\hline
p0m6  & Mean     & $-37.7\pm17.9$ & $12.0\pm19.6$  \\
      & $\sigma$ & $91.1\pm12.6$  & $94.0\pm13.9$  \\
      & Number   &     26         &    33          \\\hline
p0m8  & Mean     & $6.0\pm21.6$   & $18.5\pm18.1$  \\
      & $\sigma$ & $52.9\pm15.3$  & $51.1\pm12.8$  \\
      & Number   &     6          &     8          \\\hline
p2m9  & Mean     & $-35.7\pm27.5$ & $-18.0\pm18.7$ \\
      & $\sigma$ & $61.4\pm19.4$  & $52.8\pm13.2$  \\
      & Number   &     5          &    8           \\\hline
m1m10 & Mean     & $-10.2\pm14.7$ & $-33.6\pm13.1$ \\
      & $\sigma$ & $54.5\pm10.3$  & $49.1\pm9.3$   \\
      & Number   &    14          &     14         \\\hline
\end{tabular}
\label{tab:RC_stream_vel}
\end{table}

\begin{figure*}
 \begin{center}
 \includegraphics[width=15.3cm]{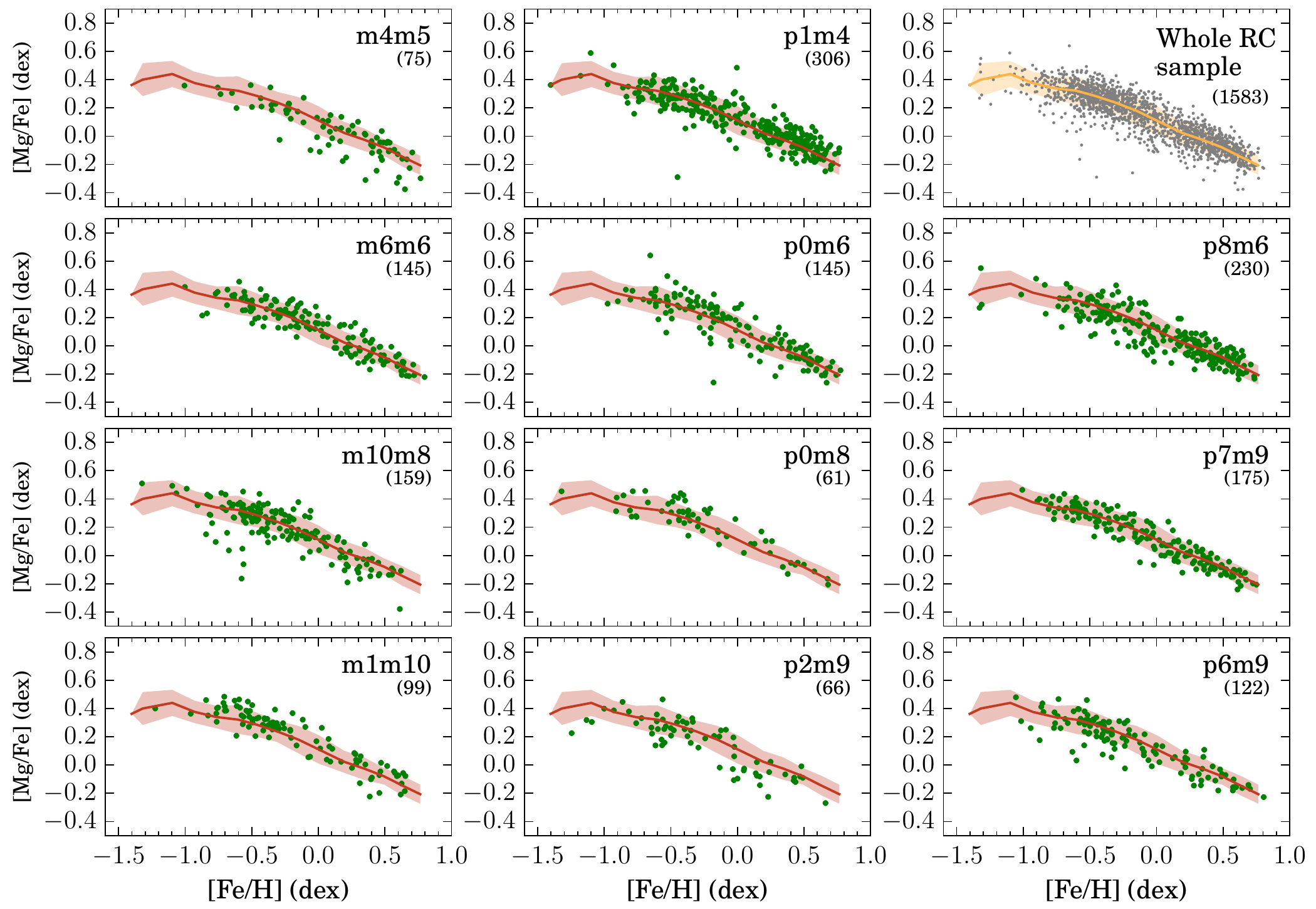}
 \caption{Sample distribution in the [Mg/Fe] vs. [Mg/Fe] plane. \textit{Upper right panel:}  Whole working sample (gray points). A fiducial median profile and $1\sigma$ dispersion band is constructed over several metallicity bins. \textit{Remaining panels:} Individual field distributions (green points) and fiducial profile and dispersion band of the whole working sample (red line and shaded area). The number of stars is given in parentheses. \textmd{The order of the panels approximately indicates the positions of fields in the ($l$,$b$) plane.}} 
 \label{fig:mgfe_vs_feh_fields}
 \end{center}
\end{figure*}

It has been suggested that an enhanced bimodality for metal-rich stars can arise or be inflated by stellar evolutionary effects \citep{nataf2014}. The RGB is redder than the RC, but both become bluer with decreasing metallicity. This implies that the relative contamination of the RC sample with RGB members can increase as a function of metallicity given a color cut in the survey selection function. From a PARSEC isochrone of 10~Gyr and $\textmd{[Fe/H]}=-1.5$~dex (so at the metal-poor end of the bulge MDF), the RC lies at $J-K=0.40$ mag, redder than the GES color cut at $J-K=0.38$ mag. Consequently, our sample should be free of this potential bias. On the other hand, the ratio of RC relative to RGB stars is an increasing function of metallicity, meaning that for example a sample with $\textmd{[M/H]}\sim-1.3$~dex should be 1.75 times larger than one at $\textmd{[M/H]}\sim0.4$ dex to display features with the same statistical significance. In the combined set of stars from the p0m6, p0m8, p2m9, and m1m10 fields, the ratio between stars with metallicity lower and higher than solar is 1.65, which ensures that this bias source might not be relevant in our case. A third potential bias comes from a metallicity dependence of the magnitude and the strength of the red giant branch bump. These factors can conspire to increase the signal of the faint magnitude peak at high metallicity. While it is true than the exact modeling of the impact of this effect is complicated, it should just increase the difference between the peaks, and does not necessarily invalidate the qualitative presence of two peaks in the magnitude distribution.

In line with previous studies in the literature \citep{DePropris2011,Uttenthaler2012,Vasquez2013}, we attempt to characterize the stream motions in the X-shape bulge by comparing the line-of-sight radial velocities of stars around the peaks of the metal-rich magnitude distribution. Given the size of our sample, this exercise may suffer from low number statistics, as evidenced by the relative size of the error bars. The results for the four studied fields are in Table~\ref{tab:RC_stream_vel}. With the exception of p0m6, there are no statistically significant differences in velocity for the bright and faint groups of metal-rich stars. These results are in agreement with previous works for p0m8 \citep{DePropris2011} and m1m10 \citep{Uttenthaler2012}. The structure of the X-shape bulge is complex; it is composed of the superposition of several stable family orbits. Radial velocity measurements on a larger number of fields might help us to unravel the nature and spatial distribution of these orbit streams.

The above analysis reinforces the bimodal nature of the MDF throughout the bulge area sampled by our fields. In the following, we aim to further characterize the MDF metallicity groups by including $\alpha$-abundances and kinematics into the analysis.

\section{Bulge trends in the [Mg/Fe] vs. [Fe/H] plane}
\label{sec:bulge_trends}

Beyond the study of the MDF, the availability of elemental abundances from high-resolution spectroscopy provides us with an important tool to understand the bulge nature. In fact, the trends displayed by stars of any stellar population in the [$\alpha$/Fe] vs. [Fe/H] plane encode important information regarding its IMF and the star formation history. This is particularly critical in Galactic bulge studies as it has been used in attempts to associate the bulge with other Galactic components, in particular with the thick disk \citep[e.g.,][]{zoccali2006,fulbright2007,alves-brito2010,bensby2013}.

The Gaia-ESO survey iDR4 provides abundances for several species. We focus here on the distribution in the [Mg/Fe] vs. [Fe/H] plane. We adopted magnesium because its abundance determination seems to be less affected by errors in stellar parameters and because its spectral lines are more clearly defined in the GIRAFFE HR21 setup domain than those of the other available $\alpha$-elements \citep{mikolaitis2014}. Moreover, like  oxygen, magnesium is expected to be produced exclusively by SN II explosions, while other alphas have more than one nucleosynthesis channel.

In Fig.~\ref{fig:mgfe_vs_feh_fields}, we display the [Mg/Fe] vs. [Fe/H] distribution of our working sample in the different fields. Here we include in p1m4 the comparison RGB and RC stars discarded while studying the MDF (since here we are interested in the trends and not in the density distribution). The upper right panel of Fig.~\ref{fig:mgfe_vs_feh_fields} shows the whole sample, together with a median profile and $1\sigma$ dispersion band calculated over several small bins in metallicity. This fiducial trend, is then overplotted on the individual field distributions in the remaining panels. The different field samples compare well with the fiducial trend; there are  no strong deviations throughout the bulge region. 

\begin{figure}
 \begin{center}
 \includegraphics[width=9.14cm]{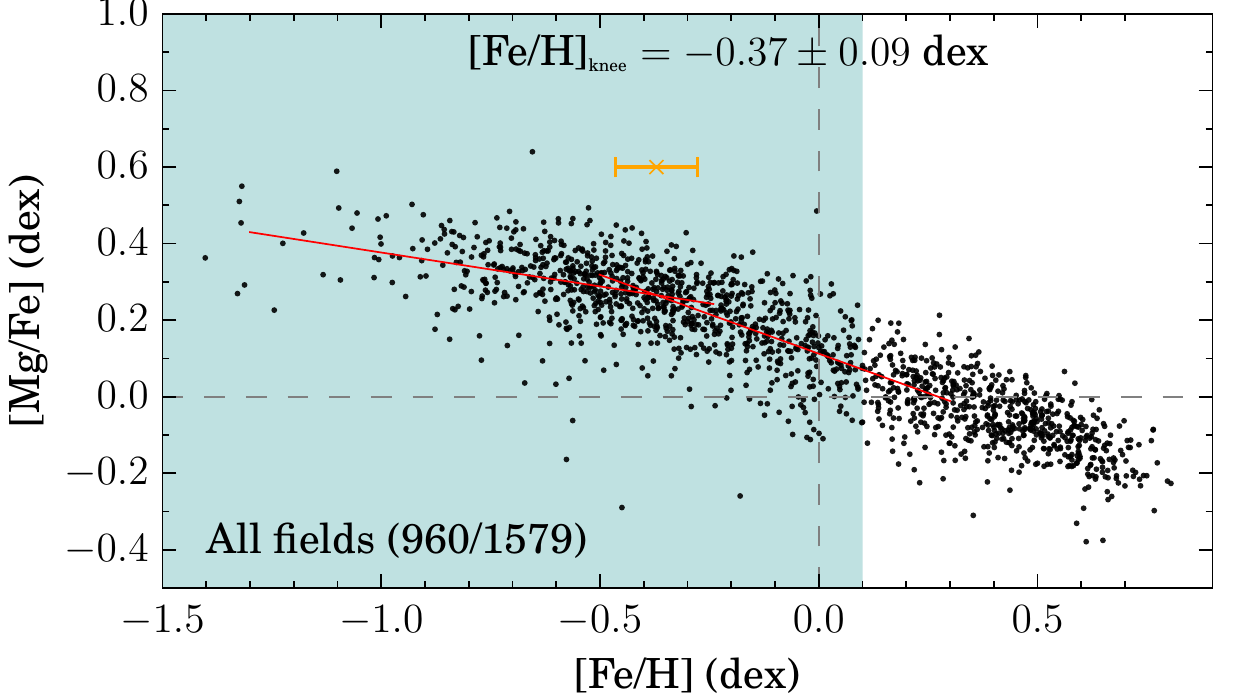}
 \caption{Determination of the bulge knee position in the [Mg/Fe] vs. [Fe/H] plane. The whole working sample is indicated by black dots. A bilinear model, fitted to the metal-poor bulge data (shaded blue area), is shown with red solid lines. The number of stars included in the fit, and the resulting knee position and error bar, are quoted in the figure. An orange error bar marks the knee position and error.} 
 \label{fig:mg_feh_slopes_bulge}
 \end{center}
\end{figure}

Moreover, Fig.~\ref{fig:mgfe_vs_feh_fields}, shows that in every field the curve tends to flatten at metallicity lower than $\sim-0.4$~dex. This is an expected feature from the time-delay model, according to which the $\alpha$-enhancement levels start to strongly decline with [Fe/H] after the maximum of the rate of supernovae Ia explosions is reached. This produces a knee in the [$\alpha$/Fe] vs. [Fe/H] trend whose location provides constraints on the formation timescale estimate of the stellar system. In Fig.~\ref{fig:mg_feh_slopes_bulge}, we present the whole bulge working sample, together with a best fit bilinear model. The model used to fit the data consists of two linear trends sharing a common point, i.e., the knee, and  leaves the other parameters free. The fit is performed in the range $-1.5\leq \textmd{[Fe/H]}\leq+0.1$, covering the metallicity range of the metal-poor bulge component. The fit is performed by means of a $\chi^2$ minimization, and errors are taken into account by performing 1000 Monte Carlo samplings from the individual errors in [Mg/Fe]. We can see in Fig.~\ref{fig:mg_feh_slopes_bulge} that given the size of the sample and the data dispersion in [Mg/Fe], we cannot constrain the knee position better than $\sim0.1$~dex, with the resulting value being $\textmd{[Fe/H]}_\textmd{knee}=-0.37\pm0.09$~dex.

The median trend of our bulge stars in the [Mg/Fe] vs. [Fe/H] plane compares  well with the distribution of inner disk stars in the [$\alpha$/Fe] vs. [Fe/H] plane presented in \citet{Hayden2015} (their Fig.~4, leftmost panels for $3<R_{GC}<5$ kpc). In both cases, the sequence starts from the locus of high-$\alpha$ metal-poor stars and ends in that of low-$\alpha$ metal-rich ones. In the case of disk stars, as seen by APOGEE, a vertical step in the sequence is visible at ${\rm[Fe/H]\sim-0.1}$~dex, which is not evident in our bulge sample. Except for this, a general similarity between the stellar distribution of bulge and disk(s) samples in the $\alpha$-abundance vs. metallicity plane can be suggested. Nevertheless, a more detailed quantitative comparison is not possible here since there is no guarantee that both surveys are in the same abundance scale. A set of common stars to cross-calibrate them is needed and awaited.

\begin{figure}
 \begin{center}
 \includegraphics[width=9cm]{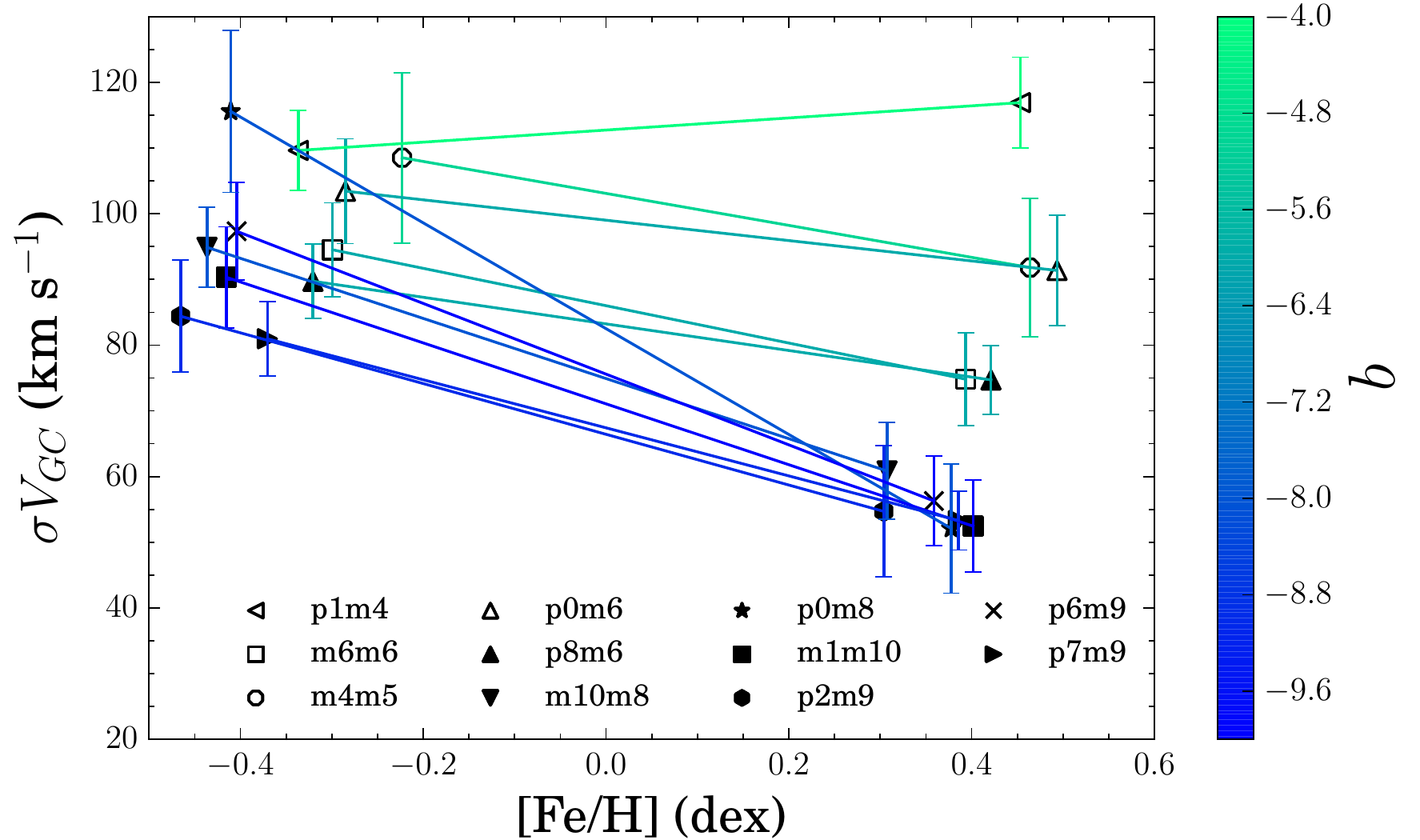}
 \caption{Velocity dispersion of metal-rich vs. metal-poor stars in each field. Points belonging to the same field  are connected by a line which is color-coded according to $b$.}
 \label{fig:vel_disp_populations}
 \end{center}
\end{figure}

Based on the conclusions drawn in Sect.~\ref{sec:MDFs} regarding the bimodal nature of the bulge MDF, we split the sample into metal-rich and metal-poor stars. To this end, we adopted the limits $\textmd{[Fe/H]}=+0.15$ and $+0.10$~dex for the fields close to ($b>7^\circ$) and far from ($b<7^\circ$)  the plane, respectively.
In Fig.~\ref{fig:vel_disp_populations}, we display the Galactocentric velocity dispersion\footnote{Galactocentric velocity conceptually corresponds to the line-of-sight radial velocity that would be observed by an stationary observer at the Sun's position. It is calculated as \begin{align*}V_{GC}&=V_{HC}+220\sin(l)\cos(b)\\&\hspace{10pt}+16.5\left[\sin(b)\sin(25+\cos(b)\cos(25)\cos(l-53)\right]).\end{align*}} trends of the fields color-coded according to their Galactic latitude \citep[for a comparison of line of sight distance distributions with simulations, see][]{williams2016}. Given that the individual radial velocity uncertainties are small compared with the field dispersions, the error in the velocity dispersion can be taken as $\sigma/\sqrt{2N}$. The bulge metal-poor components appear to be kinematically hot throughout the whole sampled area, with values around $\sigma V_{GC}=100 \textmd{km}\ \textmd{s}^{-1}$. Instead, the metal-rich components present velocity dispersions higher close to the plane, and decrease systematically with $b$.

\begin{figure}
 \begin{center}
 \includegraphics[width=8.8cm]{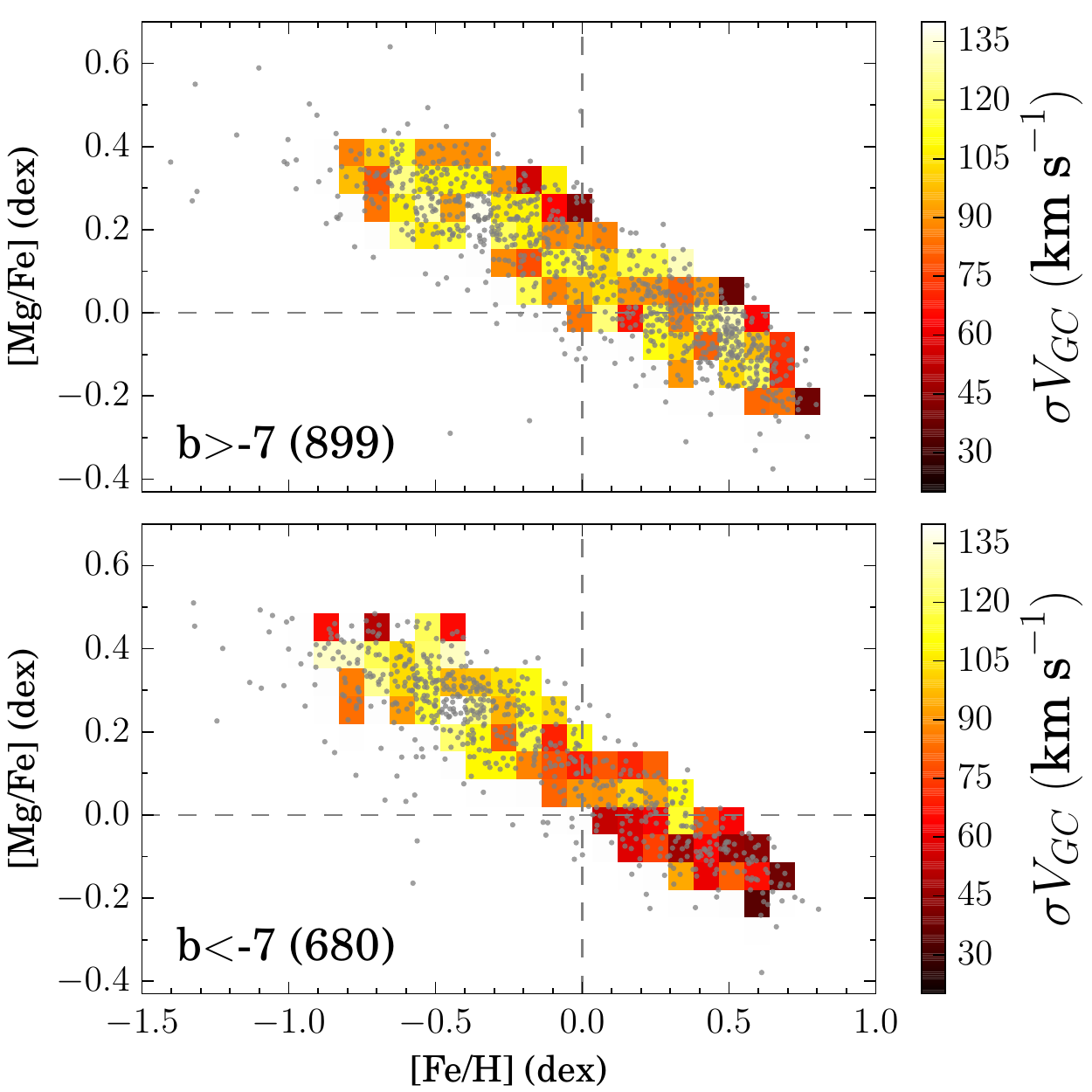}
 \caption{Velocity dispersion distribution in the [Mg/Fe] vs. [Fe/H] plane. The sample is divided into small parcels inside which velocity dispersions are calculated. They are color-coded as indicated by the colorbars. Two gray dashed lines indicate in each panel the Sun's position for reference. \textit{Upper panel:} Fields close to the plane with $b>-7$. \textit{Lower panel:} Fields far from the plane with $b<-7$. The number of stars is given in parentheses.} 
 \label{fig:vel_disp_histo2d_lats}
 \end{center}
 \end{figure}

To see these results more in perspective, we display in Fig.~\ref{fig:vel_disp_histo2d_lats} the [Mg/Fe] vs. [Fe/H] distributions of fields close to  and far from the plane. Each subsample is split into small areas, color-coded according to their velocity dispersion. On average, metal-rich and metal-poor parcels are kinematically homogeneous in inner fields, while for the outer ones the metal-rich end is  kinematically colder. It is worth noting that, according to this figure, there is no evidence of kinematic variations with [Mg/Fe] at fixed metallicity.

\begin{figure}
 \begin{center}
 \includegraphics[width=8.7cm]{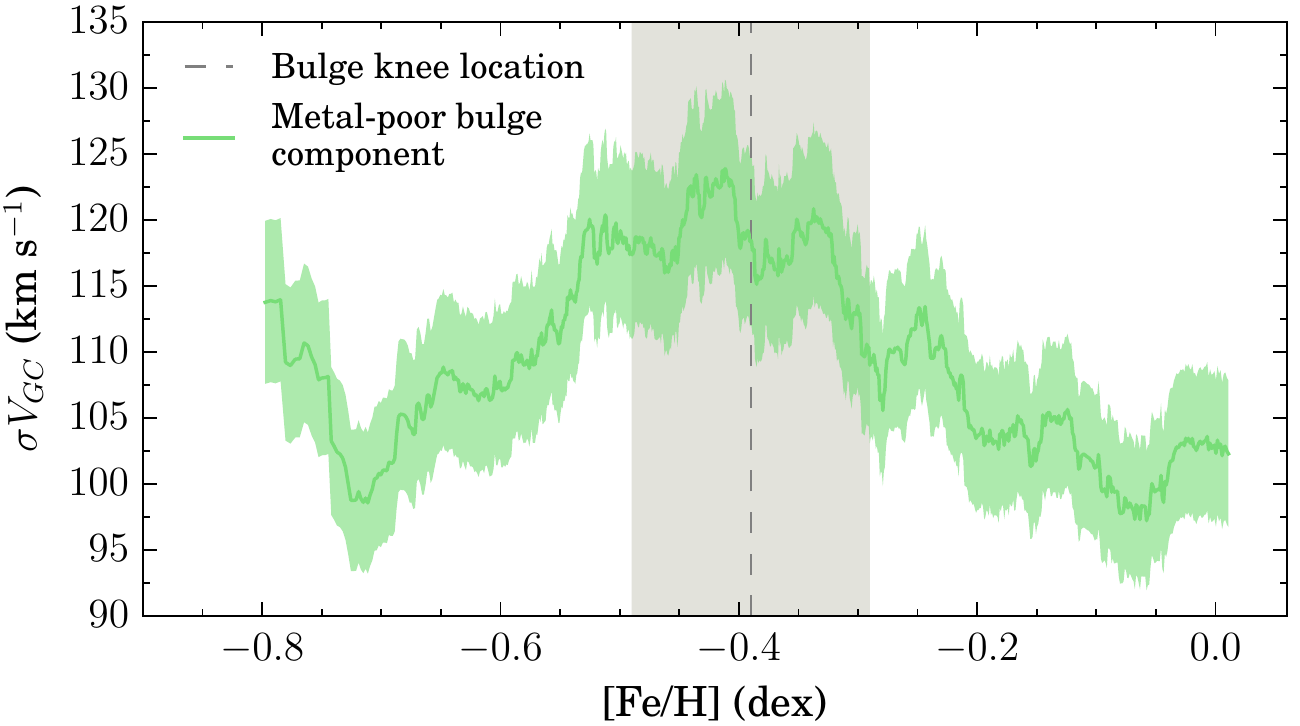}
 \caption{Velocity dispersion vs. metallicity profile of the metal-poor bulge. A running median with bin size of 170 data points is used to construct the curve displayed as a green solid line. A $1\sigma$ error band around the mean is given by the green shaded area. The metallicity and error of the bulge knee in the [Mg/Fe] vs. [Fe/H] plane are indicated by a vertical dashed line and gray shaded area.}
 \label{fig:velDisp_vs_met_mp}
 \end{center}
\end{figure}

Figures \ref{fig:vel_disp_populations} and \ref{fig:vel_disp_histo2d_lats} show that the metal-poor bulge component seems to be more kinematically homogeneous than the metal-rich one in the surveyed area. We attempt to test in detail the kinematics of the metal-poor stars by using all of them to construct the velocity dispersion profile displayed in Fig.~\ref{fig:velDisp_vs_met_mp}. A $1\sigma$ error band is displayed as a shaded area. An interesting trend is clearly visible:  the velocity dispersion increases and then decreases symmetrically around $\textmd{[Fe/H]}\sim-0.4$~dex, which -- curiously -- is  roughly the metallicity where the [Mg/Fe] vs. [Fe/H] knee is located. This is illustrated by the dashed gray line and shaded area depicting the knee's metallicity position and error. This behavior is different from that displayed, over the same metallicity range, by ARGOS data \citep[cf.][their Fig.~7; ${\rm-0.8\leq[Fe/H]\leq 0.0}$~dex]{ness2013b}. In this sense, it is not fully clear whether the velocity dispersion of metal-poor bulge stars increases steadily as a function of decreasing metallicity or presents a more complex behavior, such as that suggested by Fig.~\ref{fig:velDisp_vs_met_mp}. We expect to be able to tackle this issue with the next internal data release of the Gaia-ESO survey as its larger spatial sampling will allow us to compare trends with enough statistics at different small regions in ($l,b$). It is important to fully characterize this behavior since it might provide an important observational constraint, and an interesting fact to be explained by chemodynamical numerical models of Milky Way formation.

\section{Chemical similarities between the thick disk and the bulge}
\label{sec:chem_simil_bulge_thick_disk}
 
A major part of the Gaia-ESO survey pointings are devoted to characterizing the disk populations. We take advantage of this sample to chemically compare  the disk(s) and the bulge on the basis of a large homogeneous sample.

\begin{figure}
 \begin{center}
 \includegraphics[width=9.3cm]{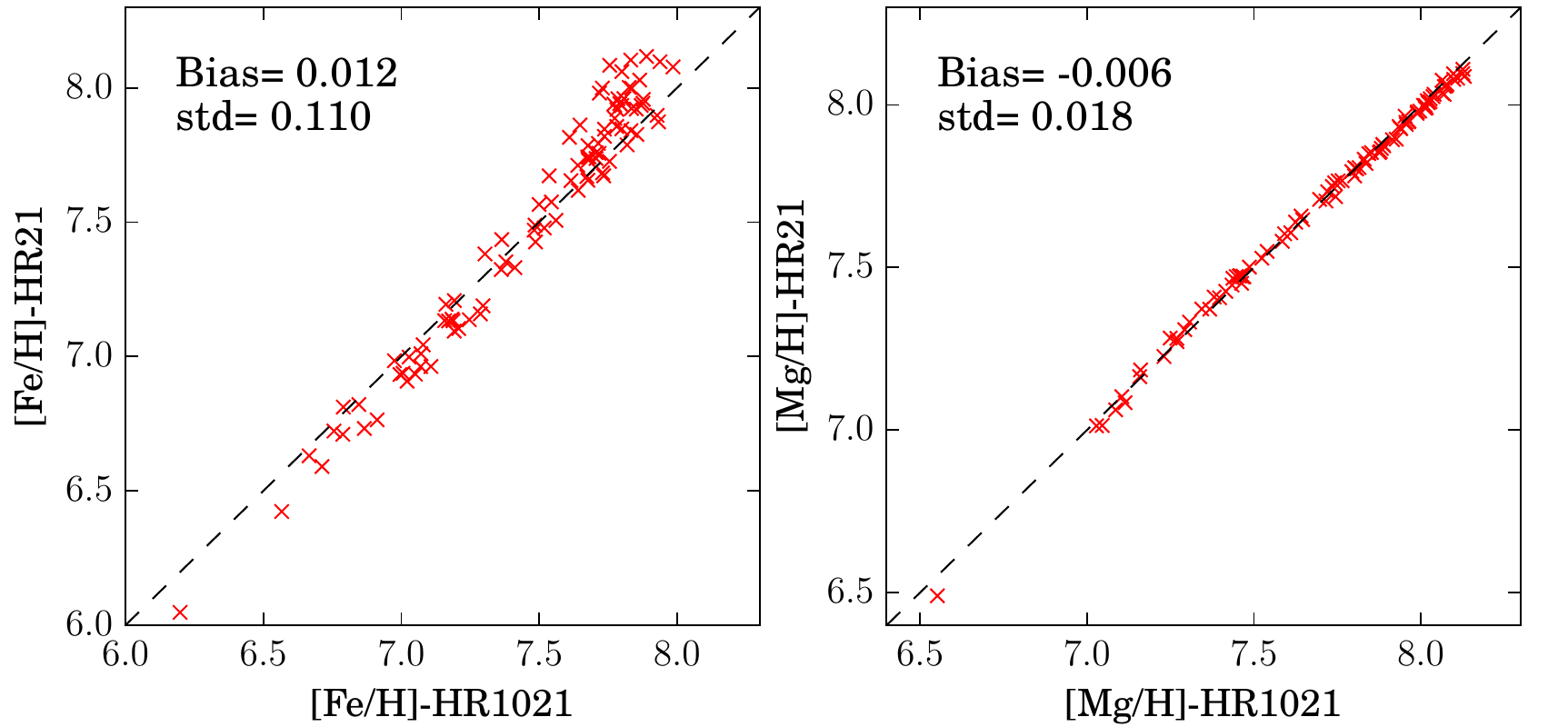}
 \caption{Iron and magnesium abundances derived from the analysis of the setup combination HR10+HR21 and HR21-only, are compared for 114 of the 228 bulge comparison stars presented in Sect. \ref{sec:datos}, which were observed in both setups.} 
 \label{fig:Fe_Mg_setup_comparisons}
 \end{center}
 \end{figure}

The disk samples of Gaia-ESO survey are observed with both the HR10 and HR21 GIRAFFE setups. A careful fundamental parameter homogenization, based on benchmark stars, ensures compatibility between the parameters and elemental abundances derived from the HR10+HR21 setup combination (disk) and the HR21 alone (bulge). In Fig.~\ref{fig:Fe_Mg_setup_comparisons} we display the comparison of HR10+HR21 and HR21 iron and magnesium abundances derived for a sample of 144 bulge stars (half of the comparison sample presented in Sect.~\ref{sec:datos}). A very good agreement between the two sets of measurements is visible.

 \begin{figure}
 \begin{center}
 \includegraphics[width=9.0cm]{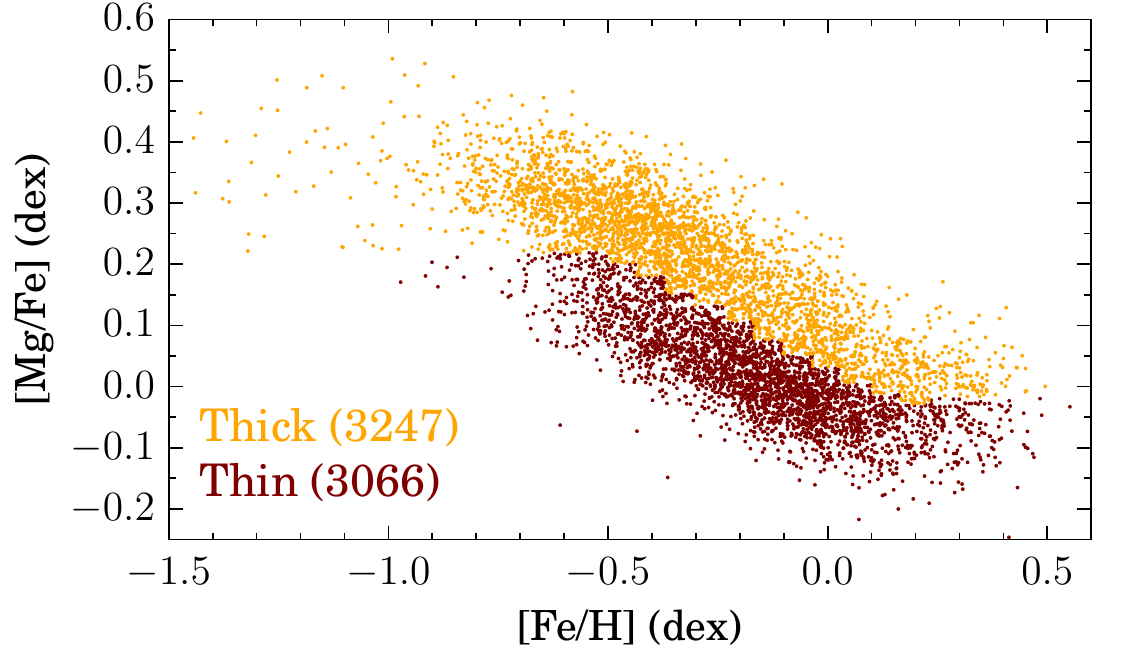}
 \caption{Selected disk subsample in the [Mg/Fe] vs. [Fe/H] plane. A sample separation into thin and thick sequences is performed as described in the main text, and color-coded;  the total number of stars in each sequence is quoted in parentheses.} 
 \label{fig:disk_sample_separation}
 \end{center}
 \end{figure}
 
  \begin{figure}
 \begin{center}
 \includegraphics[width=8.8cm]{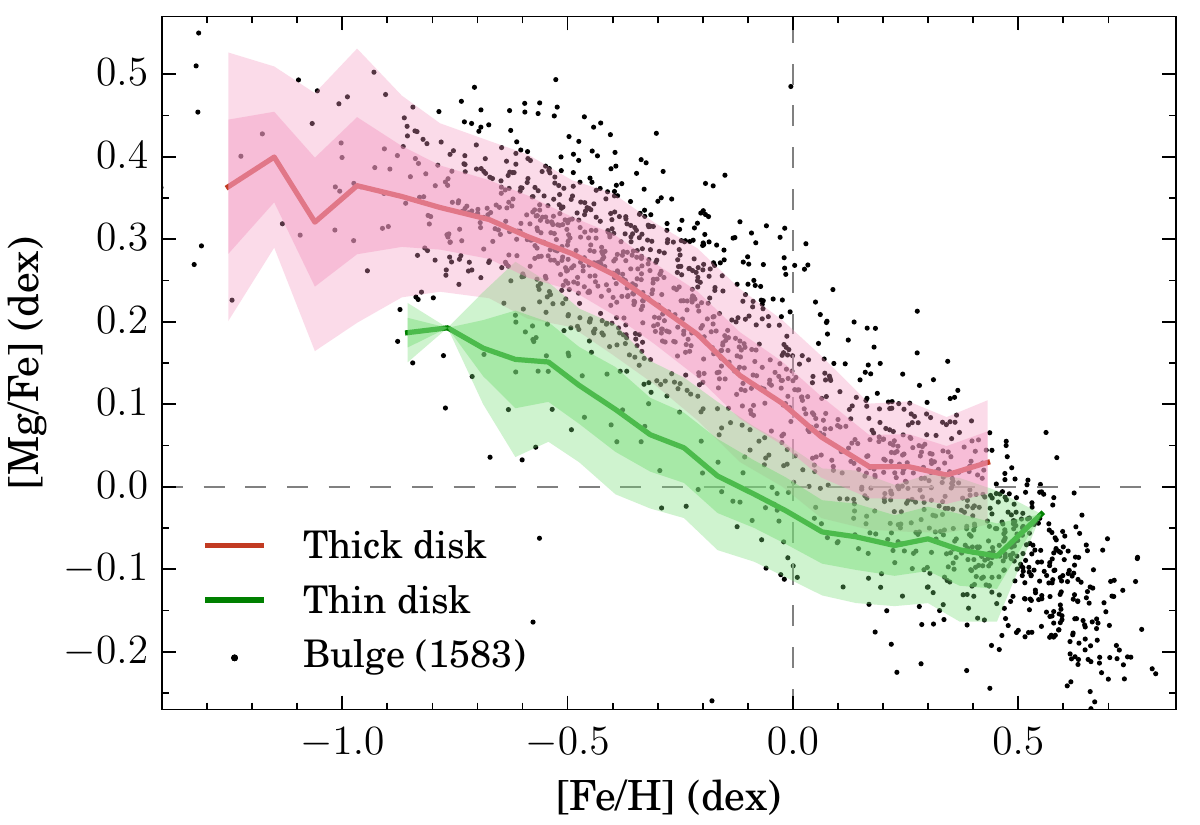}
 \caption{Bulge sample (black dots), mean trend (solid lines), and $1\sigma$ and $2\sigma$ dispersion bands (shaded areas) for the thin (green) and thick (red) disk profiles in the [Mg/Fe] vs. [Fe/H] plane.} 
 \label{fig:bulge_disk_profile_comparison}
 \end{center}
 \end{figure}

From the whole disk sample, we selected stars satisfying $\textmd{S/N}\geq45$, ${\rm\Delta T_{teff}\leq150\,K}$, $\Delta \log(g)\leq 0.23$~dex, $\Delta \textmd{[M/H]}\leq0.20$~dex, $\Delta \textmd{[Fe/H]}\leq0.1$~dex, and $\Delta \textmd{[Mg/H]}\leq0.08$~dex. In this way, we defined a clean disk sample composed of 6313 stars. A separation of thin and thick disk stars in the [Mg/Fe] vs. [Fe/H] plane was performed by following the dip in [Mg/Fe] distribution in several narrow metallicity bins. The separated subsamples are shown in Fig.~\ref{fig:disk_sample_separation}.

As a first qualitative comparison between the disks and the bulge in the [Mg/Fe] vs. [Fe/H] plane, we constructed median curves and dispersion bands for the thin and thick disk sequences. We overplotted  the resulting profiles on top of the bulge sample distribution in Fig.~\ref{fig:bulge_disk_profile_comparison}. We can see that bulge and thick disk stars have comparable [Mg/Fe] enhancement levels over the whole metallicity range spanned in common. Nevertheless, a larger dispersion in [Mg/Fe] of bulge stars relative to the thick disk is apparent along the whole metallicity range. Although this can be a real feature that  reveals differences in chemical evolution between the two populations, we cannot rule out the possibility that this effect is the result of the lack of  spectral information available from the HR21 setup for the bulge compared to the HR10+HR21 available for the thick disk sample. On the other hand, the thin disk sequence runs under the bulge one and matches it at $\textmd{[Fe/H]}>0.1$ dex. In this way,  a chemical similarity between the metal-poor bulge and the thick disk, and between the metal-rich bulge and the thin disk are apparent. This  has the important implication that if we want to explain the bulge as the product of secular evolution, we have to include both the thin disk and the thick disk to properly account for the chemical properties of the bulge sequence \citep[in line with the recent claim of][]{DiMatteo2015}. Current suggestions \citep{shen2010,martinez-valpuesta2013} include just the thin disk, which is not consistent with the chemical evidence presented here.
  
\begin{figure*}
 \begin{center}
 \includegraphics[width=14cm]{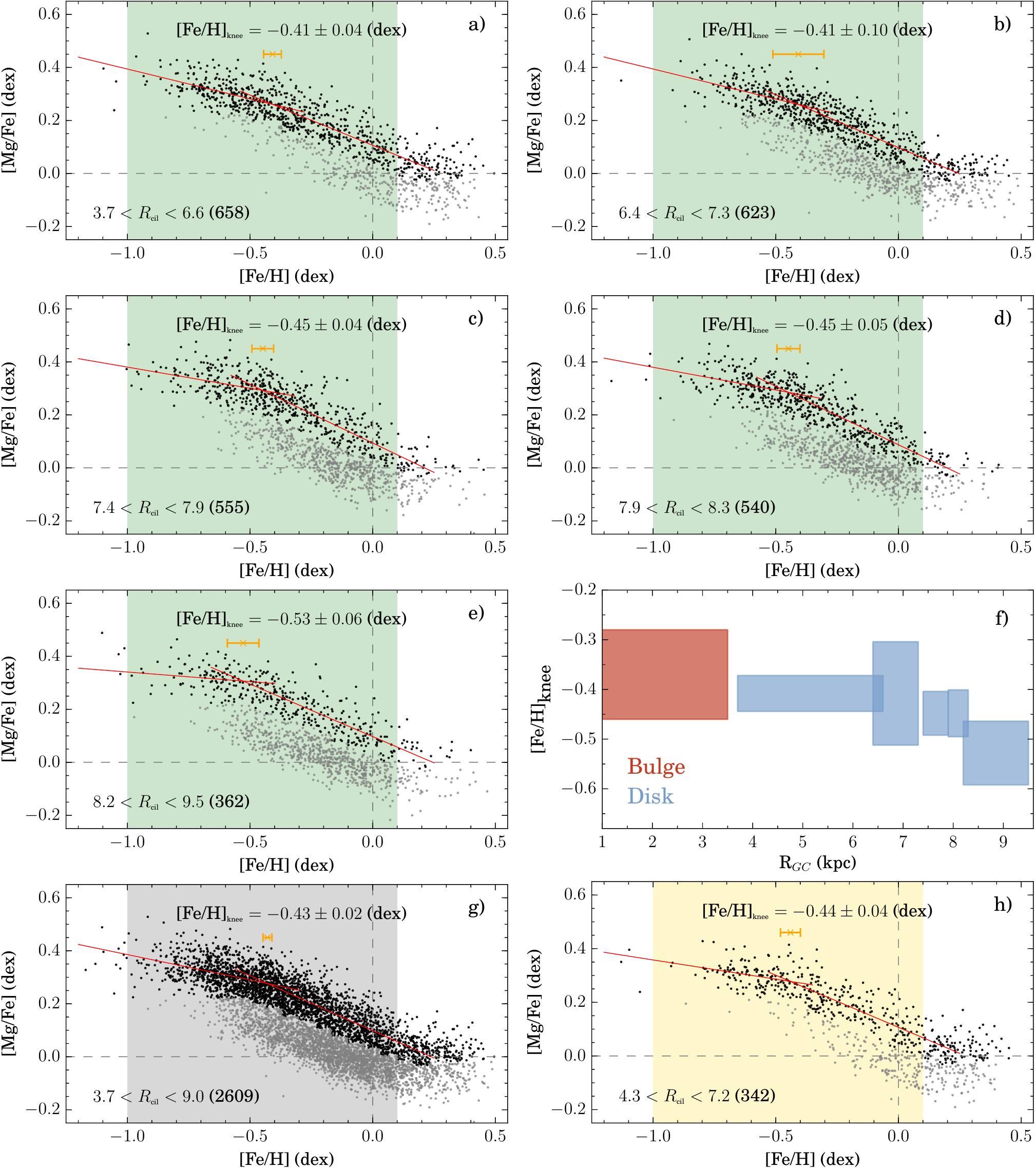} 
 \caption{ [Mg/Fe] vs. [Fe/H] distribution of disk stars in several radial distance bins with $|Z_{GC}|\leq3$ kpc. Gray and black points indicate thin and thick disk stars. A shaded area highlights the metallicity range used to perform a bilinear model fit of the thick disk sequence. The number of thick disk stars used to perform the fit is indicated in parentheses. The best fit model in each radial bin is displayed with red lines. The [Fe/H] location of the knee, together with its error bar, is quoted in each panel and is indicated by an orange error bar. \textit{Panels a-e:} Subsamples in several Galactocentric radial bins, as indicated in each panel. \textit{Panel f: } [Fe/H] position of the knee as a function of Galactocentric distance. Box length and height depicts the size of the radial bin and the error bar of the measurement. \textit{Panel g: } Whole thick disk sample, grouping together all the stars in the panels a-e. \textit{Panel h: } Subsample of RC stars in a radial range where the mean $|Z_{GC}|$ is approximately constant with $R_{GC}$.} 
 \label{fig:mg_feh_slopes_thick_disk}
 \end{center}
\end{figure*}

We attempt to make more detailed assessments of the chemical similarity between the bulge and the thick disk by comparing the metallicity location of the knee in the two sequences. Unlike previous attempts in this direction, our thick disk sample spans a broader extent in Galactocentric radii, with a significant number of stars observed down to 4 kpc. We selected stars with $|Z_{GC}|\leq3$ kpc to ensure a nearly homogeneous distribution of $Z_{GC}$ along the sampled radial range.

We split the thick disk sample in five radial portions of approximately the same number of stars in order to probe potential radial variations of the knee position. As we did for the bulge sample, we fit a bilinear model to the thick disk sequence in each radial bin. We use stars in the range $-1.0\leq \textmd{[Fe/H]}\leq+0.1$ dex to avoid the undersampled metal-poor end and the region where the thin and thick disk sequence separation is more uncertain (i.e., around solar metallicity). The results for the five radial bins are displayed in panels a-e of Fig.~\ref{fig:mg_feh_slopes_thick_disk}. The metallicity at which the knee is located, and the respective error from 1000 Monte Carlo samplings on the individual [Mg/Fe] errors, are quoted in each panel. We can see that, accounting for the error bars, the position of the thick disk knee does not change through the sampled radial region. This is explicitly shown in panel f, where -- except for the last distance bin (with lower number statistics) -- the different ${\rm [Fe/H]_{knee}}$ measurements are consistent with being flat with respect to $R_{GC}$. A radial decrease in the knee metallicity position with $R_{GC}$ would imply an inside-out formation for the thick disk, which would conflict with the observed absence of a radial metallicity gradient \citep{mikolaitis2014}. Instead, the constant ${\rm [Fe/H]_{knee}}$ we found here might imply a formation given by a single star burst in an initially well-mixed media.

A similar shape of the thick disk trend in the [$\alpha$/Fe] vs. [Fe/H] plane for all $R_{GC}$ has been also qualitatively suggested by the APOGEE data \citep{Nidever2014,Hayden2015}. As already mentioned, a quantitative detailed comparison between the GES and APOGEE results is not possible because of the unavailability of a set of common stars for cross-calibrating their abundance scales. The trends of low- and high-$\alpha$ stars displayed in Fig.~\ref{fig:mg_feh_slopes_thick_disk} and those of \citet{Hayden2015} (the middle and lower rows of their Fig.~4) are comparable:  the two disk sequences intersect each other at solar metallicity. In the inner distance bins, rather than  a single sequence of disk stars, as suggested in \citet{Hayden2015},  both sequences are visible in GES data but with a lack of metal-poor thin disk stars. This is expected if those stars constitute a different outer disk population, as has recently been  suggested \citep{Haywood2013, rojas-arriagada2016}.

Given the radial constancy of ${\rm [Fe/H]_{knee}}$ of thick disk stars, we attempt to increase the accuracy of its determination by performing a bilinear fit on the whole thick disk sample (mean Galactocentric radius $R_{GC}=7.1$~kpc). We obtained a value of $\textmd{[Fe/H]}_{knee}=-0.43\pm0.02$~dex, which we can consider as representative of the whole thick disk (panel g). In panel \textit{h}, we display a fit performed just considering RC thick disk stars. The resulting  $\textmd{[Fe/H]}_{knee}=-0.44\pm0.04$~dex is in agreement with the figure derived from the whole sample. This demonstrates that no systematics are likely to be introduced in our analysis by using results coming from the combination of dwarf and giant stars.

Finally, we compare the metallicity knee position of the thick disk and bulge sequences. A difference of $\Delta\textmd{[Fe/H]}=0.06$~dex is found. This difference is relatively small with respect to the size of the error bars of both determinations (0.02 and 0.09 dex, respectively). Unfortunately, the uncertainty levels of our abundance measurements  prevents us from making a strong assessment on the statistical significance of a null difference. However, assuming the plausible scenario of a nonzero difference, it would have an upper limit of $\Delta\textmd[Fe/H]_{knee}=0.24\ \textmd{dex}$, considering its 95\% confidence interval.

In summary, we found evidence of a constant  SFR with Galactocentric distance for the thick disk formation. In addition, a chemical similarity between the bulge and the thick disk is suggested by the data. A fine-tuned compatibility between the detailed properties of the two sequences is beyond the statistical resolution of the present sample. Nevertheless, some caution should be taken when considering the facts exposed here; although similar enhancement levels are found for the two populations, indicating a similar IMF, the bulge exhibits a larger dispersion in [Mg/Fe] around the mean, a result that needs to be confirmed with a more homogeneous data set. And similarly, although the knee metallicity positions of the two sequences are comparable within the errors, a plausible difference as large as 0.24 dex suggests a difference in the characteristic SFR of the two populations, i.e., the bulge  formed on a shorter timescale than the thick disk.

\section{Comparison with a chemical evolution model}
\label{sec:comparison_CEM}

The modeling of observational data by means of chemical evolution models provides an interesting opportunity to put constraints on the formation timescale of a stellar system. We attempt here to constrain the bulge formation timescale by adopting a model for a bulge formed at early epochs from the dissipative collapse of a cloud accompanied by a strong burst of star formation. To this end, we adopted the model of \citet{Grieco2012}. In this work, two bursts of star formation are invoked to model the metal-rich and metal-poor modes of the bulge MDF. We adopt here the model corresponding to the metal-poor bulge. The model assumes a gas infall law given by

\begin{equation}
 \left(\frac{d\sigma_\textmd{gas}}{dt}\right)_\textmd{infall}=A(r)X_\textmd{i} e^{-t/T_\textmd{inf}},
\end{equation}

where $X_\textmd{i}$ is the abundance of a generic chemical element $i$ in the infall gas, whose chemical composition  is assumed to be primordial or slightly enhanced from the halo formation; $T_\textmd{inf}$ is the infall timescale, fixed by reproducing present day abundances (MDF), SFR, and stellar mass; and $A(r)$ is a parameter fixed by reproducing the current average total bulge surface mass density. The parametrization of the star formation rate is adopted as a Schmidt-Kennicutt law:
\begin{equation}
 \psi(t)=\nu \sigma_\textmd{gas}^k
\end{equation}
with $k$ the law index and $\nu$ the star formation efficiency (i.e., the star formation rate per unit mass of gas). The model includes a Salpeter IMF, constant in space and time, which allows   the MDF of the metal-poor bulge population to be correctly reproduced. The set of yields are adopted from \citet{romano2010}.

\begin{figure}
 \begin{center}
 \includegraphics[width=9.05cm]{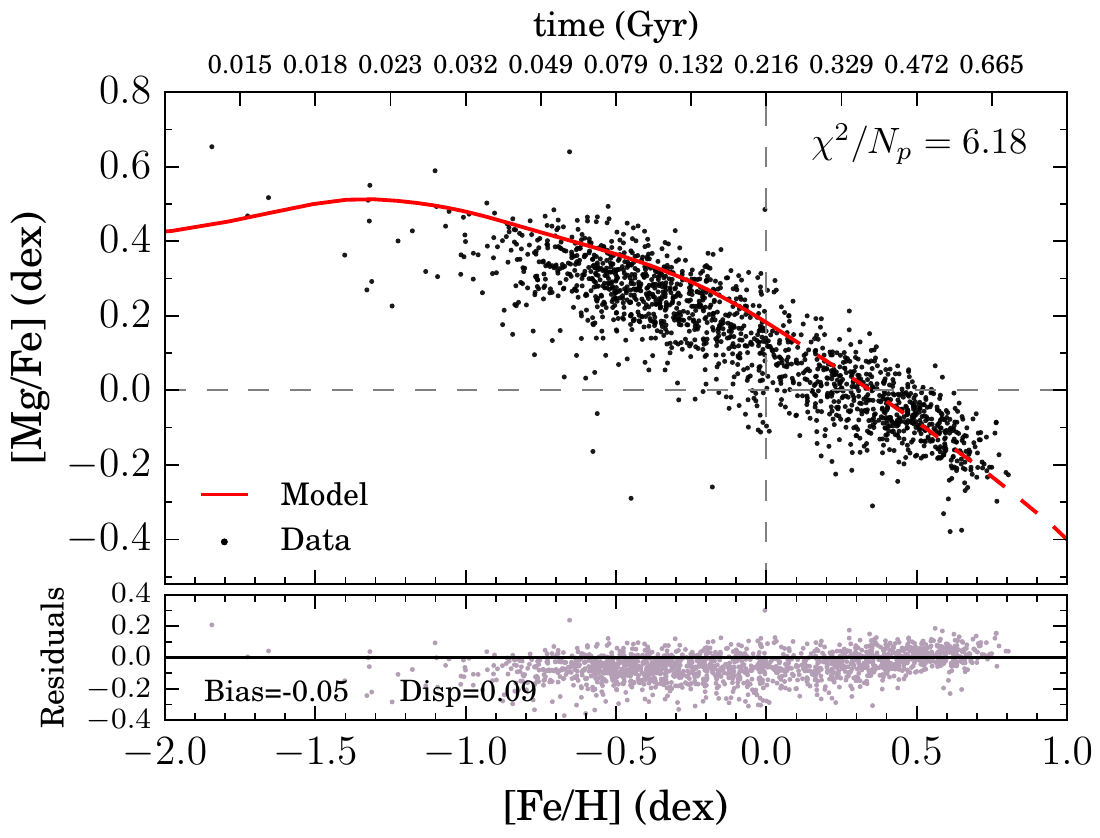}
 \caption{Comparison between the bulge data (black dots) and the predicted sequence (red line) from the chemical evolution model. The line changes from solid to dashed to emphasize that the model parameters are adjusted to fit the metal-poor bulge MDF component. \textit{Main panel: }  Evolution in time of the modeled quantities  indicated by the scale at the top of the panel. A normalized $\chi^2$ between the model and data is quoted. \textit{Small panel: }Residuals between the data and the model.} 
 \label{fig:mg_fe_salpeter}
 \end{center}
\end{figure}

We ran several models, adjusting the parameters to better reproduce the data. Our best model is displayed in Fig.~\ref{fig:mg_fe_salpeter}, where it is compared to the whole bulge working sample. This model assumes a short timescale for the gas infall $T_\textmd{inf}=0.1\ \textmd{Gyr}$ and a very efficient star formation, with $k=1$ and  $\nu=25\ \textmd{Gyr}^{-1}$.

The main characteristics of the bulge sequence (enhancement levels,  qualitative location of the knee) are well reproduced by this model. We can see that, for $\textmd{[Fe/H]}\geq-1.5$~dex, the predicted [Mg/Fe] abundance ratio steadily decreases with metallicity. This behavior increases from $\textmd{[Fe/H]}\gtrsim-0.4$~dex, which is comparable with the observed locus of the knee, as determined in Sect.~\ref{sec:bulge_trends}. The overall formation timescale, as read from the upper axis of Fig.~\ref{fig:mg_fe_salpeter}, indicates a rather rapid chemical enrichment of the bulge taking place on a timescale of 0.5-0.7~Gyr. Such a short timescale is compatible with a monolithic assembly of the metal-poor bulge. It is worth noting that the model formally shows a continuous enrichment of the gas up to super-solar metallicity. However, the number fraction of such super-solar stars is low compared to populations with lower metallicities. Thus, from the point of view of the chemical evolution model, it is not excluded that the bulge contains a fraction of metal-rich old stars born  in situ. In this sense,  metal-rich bulge stars might constitute a composite population of stars formed in situ plus a larger fraction of stars with disk origin currently located in this region as the product of secular bar-driven dynamics. This might be compatible with the relatively flat or even bimodal age distribution of  metal-rich bulge stars, as suggested from the analysis of microlensed dwarfs \citep{bensby2013} and APOGEE giants \citep{schultheis2017}, respectively.

\section{Discussion}
\label{sec:discussion}

In this work, we have made use of the fourth internal data release of the Gaia-ESO Survey to perform a fully spectroscopic analysis of the bulge in the perspective of other Galactic components. The evidence presented here leads us to consider the Galactic bulge as a composite structure, due to the coexistence of two main stellar populations.

From a Gaussian mixture models analysis, the bulge MDF appears as a bimodal distribution comprising a narrow super-solar metal-rich component  and a broad metal-poor component. This bimodal nature is verified in all the individual fields, except in those limited by small number statistics. The relative proportion of stars belonging to each  of the two populations changes, with metal-poor stars dominating far from the Galactic plane. The line-of-sight Galactocentric velocity dispersion correlates with metallicity, further stressing the likely different nature of the two populations. Metal-poor stars display a high-velocity dispersion around 100~\kms and nearly independent of ($l$, $b$). Instead, metal-rich stars present a more complex behavior;  the stars close to the plane are as kinematically hot as the bulge and decrease systematically with $b$ toward disk values. An additional correlation with metallicity appears when considering the bimodal nature of the bulge RC magnitude distribution. In the fields where this feature is visible, the distinction between the magnitude peaks is enhanced if just metal-rich stars are considered, while metal-poor ones display flatter distributions.

This bimodality contrasts with the trimodal ARGOS MDFs found by \citet{ness2013}. Their distributions are derived from a larger sample of stars ($\sim10200$), grouped together  in three $l=\pm15^\circ$ latitude strips. If the metal-rich and metal-poor bulge populations have some intrinsic gradients with spatial location, like those reported here, a MDF assembled from fields spanning a large region might have components slightly smeared out in metallicity, thus displaying a more complex MDF. Upon completion of the next GES releases, including a larger number of observed fields, we will be able to test this possibility further. In addition, the ARGOS analysis (based on data of lower resolution and mean S/N than our GES data) is not fully spectroscopic since it made use of photometric constraints to estimate fundamental parameters and metallicity. A number of recent studies examining specific locations in the bulge region \citep{Uttenthaler2012, gonzalez2015,schultheis2017} find a bimodal MDF. Instead, ARGOS MDFs are trimodal even in individual fields, where the sample size is comparable to that of the above-mentioned studies. In this sense, as discussed in \citet{schultheis2017}, it might be a possibility that the trimodal nature  of the ARGOS MDFs is a feature that arises from their parametrization and the spatial distribution of their fields rather than as an effect of larger number statistics.

Two different origins can be proposed for the stars belonging to the metal-rich and metal-poor MDF components. Metal-rich RC stars participate in the B/P bulge, present bar-like kinematics, and are chemically comparable to metal-rich thin disk stars. We associate them, in agreement with the literature, with a population formed by the classical mechanism of secular evolution of the disk via bar formation and buckling into an X-shape structure \citep{combes1981,raha1991,Athanassoula2005, martinez-valpuesta2006}. An internal vertical metallicity gradient, like the one reported here, is predicted by N-body simulations of secular bulge formation as an effect of the mapping in the vertical direction of horizontal \citep{martinez-valpuesta2013} or vertical \citep{bekki2011} metallicity gradients initially present in the disk, or a combination of the two \citep{DiMatteo2015}. Thin disk stars in the bulge region can  explain the presence of young stars \citep{bensby2013}, and also the existence of the thin star forming inner disk in the bulge traced by classical Cepheids identified from VVV photometry \citep{dekany2015}.

On the other hand, metal-poor RC stars do not participate in the B/P bulge, dominate in number density far from the plane, and display isotropic kinematics. These stars might be associated with a classical spheroid component formed at early times from the dissipative collapse of a primordial cloud accompanied by a strong burst of stellar formation. A radial internal metallicity gradient, like the one reported here, and a high-velocity dispersion are expected features from a dissipative collapse. The enhanced levels of $\alpha$-elements and the metallicity at which the knee of the sequence in the [Mg/Fe] vs. [Fe/H] plane takes place are both interpreted here with a chemical evolution model as signatures of a fast chemical evolution ($t\leq1$ Gyr), dominated by massive stars and characterized by a high star formation efficiency.
\citet{shen2010} suitably reproduced the bulge rotation and dispersion curves from the BRAVA project with an N-body simulation which limits the mass contribution of a possible dissipative collapse-made bulge to be less than 8\%. This scenario is incompatible with our results. The mean proportion of stars belonging to each of the MDF components is weighted toward metal-poor stars. Although our data set may not be fully adequate to make strong assessments on the mass contribution of both populations to the global bulge mass budget (the sample is not large enough and there is not enough spatial coverage, especially in the inner bulge), the data clearly hints at a bulge composed of a similar fraction of metal-poor and metal-rich stars. Moreover, from the observed mass-metallicity relation for galactic spheroids \citep{gallazzi2005}, a system with mean metallicity around ${\rm [Fe/H]=-0.4\,dex}$, as our metal-poor component is, should have a mass of $\sim10^{10}\,M_\odot$, which is comparable to the total bulge mass as estimated from observations \citep{valenti2016}. Furthermore, RR Lyrae stars, as tracers of metal-poor old stellar populations, have been shown to display an axisymmetric spatial distribution, uncorrelated with the bar position angle \citep{dekany2013, Gran2016}, and a high-velocity dispersion of around 130~\kms \citep{Gratton1987,Kunder2016}. However, this is still under debate; a recent analysis of OGLE data suggests that the RR Lyrae distribution might be elongated with a pivot angle comparable to that of the main bar \citep{pietrukowicz2015}. The fact that metal-poor bulge stars present cylindrical rotation has been taken as an argument for their secular origin \citep{ness2013b}. However, recent N-body simulations have shown that an initially nonrotating classical bulge can  spin-up into a bar-like structure by absorbing a significant fraction of the disk angular momenta emitted by the bar during its secular evolution \citep{saha2012}. In this way, the classical bulge will become photometrically and kinematically indistinguishable from the B/P bulge.  In the case of a massive initial classical bulge, its central parts might be less affected \citep{saha2016}, providing a kinematic relic to be exploited by spectroscopic observations of the inner-bulge.

Despite the above discussion, the comparable $\alpha$-enhancement levels, along with the similar position of the knee of the bulge and thick disk sequences, argues for a possible common origin, or at least for a similar chemodynamical evolution of these populations. Moreover, it has been shown that it is possible to reproduce general kinematic and chemical bulge patterns from N-body models explaining its formation as the product of thin+thick disk evolution \citep{DiMatteo2015}. For the knee, we found a small difference of 0.06 dex and an upper limit of 0.24~dex. In the same vein, although similar $\alpha$-enhancement levels are found for the bulge and the thick disk, there is an indication of a larger dispersion in the bulge than in the thick disk. Moreover, enhancement differences in $r$- and $s$-process elemental abundances have been proposed as evidence of different formation timescales between the two structures \citep{Johnson2012,VanDerSwaelmen2016}. Unfortunately, current efforts in this direction are based on the comparison of bulge giants and local dwarf samples, which might suffer from systematics given the different temperature and gravity regimes of the samples. All things considered, the origin and nature of the metal-poor bulge remains to be firmly defined on the basis of larger data sets and further detailed modeling.

The composite nature of galaxy bulges has been pointed out by several authors,  from a theoretical point of view  \citep{Samland2003,Athanassoula2005,Obreja2013,Fiacconi2015} and  from observational evidence in external galaxies \citep{Gadotti2009,Nowak2010,Williams2011,Erwin2015,Fisher2016}. Rather than an eccentricity of nature, the presence of composite bulges, with two or more structure types  (disk, pseudobulge, classical bulge) coexisting in the same galaxy, appears as a common outcome of galaxy formation and evolution.

If we assume that the differences in the knee position and [Mg/Fe] dispersion between the bulge and thick disk sequences are real, we can draw a general picture of the bulge formation by interpreting the observational evidence in terms of two different formation episodes. On the one hand,  the  old bulge population that formed in  situ is the product of a fast dissipative collapse in the early epochs of Milky Way evolution. As characterized by a strong SFR, the chemical enrichment of the gas may have reached super-solar metallicities before gas exhaustion, with the majority of stars produced around ${\rm [Fe/H]=-0.5\,dex}$. On the other hand, metal-rich stars in the X-shaped bulge are, as pointed out in the literature, the product of the secular evolution of the early inner disk. In this sense, the main epoch of chemical enrichment in the inner Galaxy occurred early, before the formation of the B/P bulge. A small fraction of the metal-rich stars, endemic to the central regions of the Galaxy, might be old, being currently outnumbered by stars with their origin in the early disk.

A semantic issue is then raised. In fact, we legitimately call ``bulge'' all stellar populations currently present in the central kiloparsecs of the Milky Way regardless of their origin and specific evolutionary histories.

The Gaia-ESO survey multi-method, model-driven, fully spectroscopic analysis of high-resolution high-S/N data provides a homogeneous self-consistent account of the main Galactic components. Using this exquisite data set, this is the first time that the bulge MDF has been characterized in a large spatial area, in individual fields, from a fully spectroscopic analysis. This is also the first attempt to compare the metallicity position of the bulge and thick disk knee based on a statistically significant sample of homogeneously analyzed stars. All in all, a composite picture of the Galactic bulge can be unambiguously established, with all the presented evidence pointing to the presence of two main components currently coexisting in the central regions of the Milky Way.

\begin{acknowledgements}
This work was partly supported by the European Union FP7 programme through ERC grant number 320360 and by the Leverhulme Trust through grant RPG-2012-541. We acknowledge the support from INAF and the Ministero dell'Istruzione, dell'Universit\`a e della Ricerca (MIUR) in the form of the grant ``Premiale VLT 2012''. The results presented here benefited from discussions held during the Gaia-ESO workshops and conferences supported by the ESF (European Science Foundation) through the GREAT Research Network Programme. A. Recio-Blanco, P. de Laverny, and V. Hill acknowledge the Programme National de Cosmologie et Galaxies (PNCG) of CNRS/INSU, France, for financial support. A. Recio-Blanco, P. de Laverny, and V. Hill acknowledge financial support form the ANR 14-CE33-014-01. M. Zoccali gratefully acknowledges support from the Ministry of Economy, Development, and Tourism's Millenium Science Initiative through grant IC120009, awarded to the Millenium Institute of Astrophysics (MAS), by Fondecyt Regular 1150345 and by the BASAL-CATA Center for Astrophysics and Associated Technologies PFB-06.

\end{acknowledgements}

\bibliographystyle{aa}
\bibliography{biblio}

\begin{appendix}

\section{GMM output parameters}
Individual GMM decompositions were performed on the 11 bulge fields studied in this paper. All except two MDFs were found to be better explained by a model consisting of two Gaussians, according to the Akaike Information Criterion. In Table. \ref{tab:gmm_parameters}, we present the set of parameters characterizing these optimal models for comparison with future studies.

\begin{table*} 
\centering
\caption{Parameters of the best GMM model for each field where two components were found. The subscripts ``MP'' and ``MR''  refer to the metal-poor and metal-rich Gaussian components, respectively.}
\begin{tabular}{l|ccc|ccc}
\hline
\hline
Field  & $\overline{\textmd{[Fe/H]}}_{MP}$ & $\sigma_{MP}$ & Weight$_{MP}$ & $\overline{\textmd{[Fe/H]}}_{MR}$ & $\sigma_{MR}$ & Weight$_{MR}$    \\
name       &     &     &                          &                &                 &                  \\\hline
m4m5  &   -0.09  &    0.36  &    0.65 &    0.51 &     0.13 &     0.35  \\
p1m4  &  -0.33   &   0.31   &   0.64  &   0.43  &    0.15  &    0.36  \\
m6m6  &  -0.30   &   0.27   &   0.60  &   0.40  &    0.20  &    0.40  \\
p0m6  &  -0.27   &   0.31   &   0.64  &   0.49  &    0.15  &    0.36  \\
p8m6  &  -0.25   &   0.35   &   0.63  &   0.44  &    0.15  &    0.37  \\
m10m8 &   -0.40  &    0.29  &    0.78 &    0.33 &     0.17 &     0.22  \\
p7m9  &  -0.44   &   0.25   &   0.49  &   0.30  &    0.23  &    0.51  \\
m1m10 &   -0.44  &    0.26  &    0.68 &    0.39 &     0.19 &     0.32  \\
p6m10 &   -0.43  &    0.26  &    0.70 &    0.38 &     0.20 &     0.30  \\\hline
\end{tabular}
\label{tab:gmm_parameters}
\end{table*}
  
 \end{appendix}

\end{document}